\documentclass[trackchanges]{aastex701}
\usepackage{longtable}
\usepackage{gensymb}
\usepackage{amsmath}
\usepackage{natbib}
\usepackage[utf8]{inputenc}

\DeclareUnicodeCharacter{03C3}{$\sigma$}
\DeclareUnicodeCharacter{200B}{\hspace{0pt}}
\DeclareUnicodeCharacter{00B1}{\pm}
\DeclareUnicodeCharacter{2212}{-} 
\shorttitle{Transit Timing of Exoplanet WASP-12 b}
\shortauthors{Biswas et al.}

\begin{document}

\title{Revisiting the Orbital Dynamics of the Hot Jupiter WASP--12\,b with New Transit Times}

\author[0009-0003-8446-4557]{Shraddha Biswas}
\affiliation{Indian Centre For Space Physics \\
466, Barakhola, Singabari road, Netai Nagar, Kolkata, West Bengal, 700099}
\email[show]{hiyabiswas12@gmail.com}
 
\author[0000-0001-7359-3300]{Ing-Guey Jiang}
\affiliation{Department of Physics and Institute of Astronomy \\
National Tsing Hua University, Hsinchu 30013, Taiwan}
\email[show]{jiang@phys.nthu.edu.tw}

\author[0000-0001-8677-0521]{Li-Chin Yeh}
\affiliation{Institute of Computational and Modeling Science \\
National Tsing Hua University, Hsinchu 30013, Taiwan}
\email[show]{lichinyeh@mx.nthu.edu.tw}

\author{Hsin-Min Liu}
\affiliation{Department of Physics and Institute of Astronomy \\
National Tsing Hua University, Hsinchu 30013, Taiwan}
\email{shayna501@gapp.nthu.edu.tw}

\author{Kaviya Parthasarathy}
\affiliation{Department of Physics and Institute of Astronomy \\
National Tsing Hua University, Hsinchu 30013, Taiwan}
\email{kaviyasarathy1998@gmail.com}

\author[0000-0002-8988-8434]{D. Bisht}
\affiliation{Indian Centre For Space Physics \\
466, Barakhola, Singabari road, Netai Nagar, Kolkata, West Bengal, 700099}
\email[show]{devendrabisht297@gmail.com}

\author[0000-0002-0193-1136]{Sandip K Chakrabarti}
\affiliation{Indian Centre For Space Physics \\
466, Barakhola, Singabari road, Netai Nagar, Kolkata, West Bengal, 700099}
\email{sandipchakrabarti9@gmail.com}

\author {D. Bhowmick}
\affiliation{Indian Centre For Space Physics \\
466, Barakhola, Singabari road, Netai Nagar, Kolkata, West Bengal, 700099}
\email{debashisbhowmick@gmail.com}

\author {Mohit Singh Bisht}
\affiliation{Indian Centre For Space Physics \\
466, Barakhola, Singabari road, Netai Nagar, Kolkata, West Bengal, 700099}
\email{mohitsinghbisht724@gmail.com}

\author{A. Raj}
\altaffiliation{Uttar Pradesh State Institute of Forensic Science (UPSIFS)}
\affiliation{Indian Centre For Space Physics \\
466, Barakhola, Singabari road, Netai Nagar, Kolkata, West Bengal, 700099}
\email{ashishpink@gmail.com}

\author{Bryan E. Martin}
\affiliation{Utah Desert Remote Observatory, Beryl, Utah}
\email{tigerbutteobservatory@gmail.com}

\author{R. K. S. Yadav}
\affiliation{Aryabhatta Research Institute of Observational Science\\
Manora Peak, Nainital 263002, India}
\email{rkant@aries.res.in}

\author{Geeta Rangwal}
\affiliation{Aryabhatta Research Institute of Observational Science\\
Manora Peak, Nainital 263002, India}
\email{geetarangwal91@gmail.com}

\begin{abstract}

In this study, we examine the transit timing deviations of the extensively studied hot Jupiter WASP-12\,b using a comprehensive dataset of 391 transit light curves. The dataset includes 7 new photometric observations obtained with the 1.3 m Devasthal Fast Optical Telescope, the 0.61 m VASISTHA telescope, and the 0.3 m AG Optical IDK telescope, along with 119 light curves from the Transiting Exoplanet Survey Satellite (TESS), 97 from the Exoplanet Transit Database (ETD), 34 from the ExoClock Project, and 134 from previously published sources. To ensure homogeneity and precision, we modeled all 391 light curves and determined their mid-transit times. A detailed transit timing analysis revealed a significant orbital decay rate of \textbf{$-31.97 \pm 0.80~\mathrm{ms~yr^{-1}}$}, corresponding to a stellar tidal quality factor of $Q'_\star = (1.52 \pm 0.038) \times 10^{5}$, thereby confirming that the orbit of WASP-12\,b is indeed decaying rapidly. Furthermore, the computation of model selection metrics ($\chi^2_r$, BIC, AIC) favors orbital decay as the most likely explanation. However, the presence of an eccentricity above the threshold value allows apsidal precession to remain a viable alternative. We also derived a planetary Love number of $k_p = 0.63 \pm 0.089$, consistent with Jupiter's value, suggesting a similar internal density distribution. In this study, orbital decay is strongly supported, as a plausible cause of the timing deviations observed in WASP-12 system. Continued high-precision monitoring will be essential to further constrain the system’s orbital evolution.

\end{abstract}




\section{Introduction}
\label{sec:intro}
The discovery of 51 Pegasi\,b by Mayor and Queloz in 1995 (\citealt{1995Natur.378..355M}) marked a milestone in exoplanet research, signaling a new era in the study of planets beyond our Solar System. Since this discovery, hot Jupiters, which are gas giant planets with masses similar to Jupiter with short orbital periods ($P < 10\,~$days), have attracted considerable attention and are now among the most extensively characterized exoplanets. For a planet following Keplerian orbit, transits are expected to occur at strictly regular intervals, reflecting a constant orbital period. Deviations from this strict periodicity can indicate the presence of dynamical interactions, such as gravitational perturbations from additional bodies, tidal effects, or relativistic influences (\citealt{2005MNRAS.359..567A, 2005Sci...307.1288H}). In particular, tidal interactions in close-in hot Jupiter systems can produce measurable variations in orbital periods (\citealt{2009ApJ...692L...9L}).

Hot Jupiters are especially prone to tidal interactions with their host stars due to their large masses and small orbital separations. Such interactions are now recognized as a key driver of planetary orbital evolution. Through tidal dissipation, i.e., the conversion of tidal forces into internal friction within the stellar interior (\citealt{1980A&A....92..167H}), these interactions can circularize planetary orbits, enforce spin–orbit synchronization, and, in systems where the planet orbits faster than the stellar rotation rate, trigger tidally driven orbital decay. Previous studies have highlighted the role of stellar age and tidal interactions in shaping the fate of close-in gas-giant planets. For instance, \citet{2023AJ....166..209M} showed that the occurrence rate of hot Jupiters around Sun-like stars declines with stellar age, implying that close-in gas giants are gradually removed over time, likely due to tidal interactions with their host stars. Similarly, \citet{2019AJ....158..190H} argued that tidal interactions can lead to the destruction of hot Jupiters while their host stars are still on the main sequence. These results provide a strong motivation to investigate orbital decay in close-in exoplanetary systems, as it may play a key role in explaining the observed distribution of hot Jupiters. Tidal dissipation is commonly parameterized by the modified stellar tidal quality factor, $Q_{\ast}^{\prime}$. Within this context, a subclass of hot Jupiters with day side temperatures exceeding ~2200 K and orbital periods of $\lesssim 2 \ \text{days}$, known as ultra-hot Jupiters (UHJs; \citealt{2018A&A...617A.110P}), has emerged as particularly favorable systems for exhibiting strong tidal effects.

With observational baselines for many systems now extending beyond a decade, signatures of orbital decay have been reported in an increasing number of hot Jupiters. Candidate systems include HAT-P-19\,b (\citealt{2022AJ....164..220H}), HAT-P-32\,b (\citealt{2022AJ....164..220H}), HAT-P-51\,b (\citealt{2024NewA..10602130Y}), HAT-P-53\,b (\citealt{2024NewA..10602130Y}), KELT-9\,b (\citealt{2023A&A...669A.124H}), TrES-1\,b (\citealt{2022AJ....164..220H, 2022ApJS..259...62I}), TrES-2\,b (\citealt{2022AJ....164..220H, 2024AJ....168..176B}), and TrES-3\,b, although the evidence remains marginal (\citealt{2022AJ....164..220H}). For TrES-5\,b, transit-timing variations consistent with nonlinearity, though not necessarily decay, have been reported (\citealt{2021A&A...656A..88M, 2022AJ....164..220H, 2022ApJS..259...62I, 2024NewA..10602130Y, 2025PSJ.....6..292R}). Compelling evidence for orbital decay has been reported for WASP-4\,b (\citealt{2020ApJ...893L..29B, 2022AJ....164..220H, 2023A&A...669A.124H}). More recently, \citet{2025MNRAS.541..714B} identified a decreasing orbital period in WASP-4\,b, further strengthening the case for tidal orbital decay in close-in gas giants; however, \citet{2025PSJ.....6..300W} found that the observed period variations are more consistently explained by light-travel time effects (LTTE) rather than true orbital decay. For WASP-32\,b, only weak evidence for orbital decay has been reported (\citealt{2023MNRAS.520.1642S}). The status of WASP-43\,b remains uncertain, with orbital decay claimed by \citet{2023MNRAS.520.1642S} but not confirmed by \citet{2022AJ....164..220H}. XO-3\,b has also been proposed as a potential orbital-decay candidate (\citealt{2022ApJS..259...62I, 2022PASP..134b4401Y}).

To date, the most compelling observational evidence for orbital decay has been obtained for the ultra-hot Jupiter WASP-12\,b (\citealt{2017AJ....154....4P, 2019MNRAS.490.1294B, 2020ApJ...888L...5Y}). The reality of orbital decay in WASP-12\,b has been further confirmed by \citet{2025PSJ.....6..300W}. Moreover, high-resolution imaging has revealed that WASP-12 is part of a hierarchical triple stellar system, with two distant M-dwarf companions physically bound to the planet-host star \citep{2014ApJ...788....2B}. Such higher-order stellar architectures can influence the long-term dynamical and tidal evolution of close-in planets, providing important context for interpreting transit timing variations and orbital decay signals. Motivated by these results, we focus our study on WASP-12\,b, an inflated ultra-hot Jupiter discovered by \citet{2009ApJ...693.1920H} as part of the Wide Angle Search for Planets (WASP) project. WASP-12\,b has a mass of $1.47 \pm 0.07 , M_{\mathrm{J}}$ and a radius of $1.90 \pm 0.06 , R_{\mathrm{J}}$ (\citealt{2017AJ....153...78C}), and it orbits a late-F type main-sequence star of $\sim 1.4 , M_{\odot}$ (\citealt{2017AJ....153...78C}) with a period of $\sim 1.09 , \mathrm{days}$. Since its discovery, it has become one of the most intensively studied UHJs, owing to its extreme physical characteristics and its unique role as the first exoplanet with confirmed orbital decay. The primary aim of this work is to present a new transit-timing analysis of WASP-12\,b by incorporating additional transit observations, thereby extending the temporal baseline available for detecting orbital evolution.

The detection of orbital decay has been enabled by combining long-term transit-timing data from both ground- and space-based surveys. These coordinated efforts have extended the observational timelines of transiting exoplanets, particularly hot Jupiters, over multiple decades, allowing the identification of subtle variations in orbital periods. A major contribution in this area has been made by \citet{2022ApJS..259...62I} (hereafter IW22), who compiled a comprehensive catalog of transit timing measurements, including both literature data and observations from the Transiting Exoplanet Survey Satellite. Their database covers 348 systems, of which 240 are hot Jupiters, and the addition of TESS data revealed tentative orbital decay in nine systems not previously identified. Since its launch in 2018, TESS (\citealt{2014SPIE.9143E..20R}) has played a central role in exoplanetary science, providing high-precision transit timing measurements for nearly all known hot Jupiters orbiting stars brighter than approximately 13th magnitude (\citealt{2022AJ....163...79H}). This capability has enabled systematic searches for orbital decay across the population of close-in gas giants.

In this study, we re-processed TESS transit timings and incorporated newly observed transits from multiple ground-based observatories worldwide, together with previously published light curves, to extend the temporal baseline of WASP-12\,b. To further enhance coverage, we also included publicly available data from the Exoplanet Transit Database and the ExoClock Project, which compile contributions from both professional and amateur astronomers. Such combined efforts in regular transit monitoring provide one of the most effective approaches for probing long-term orbital variations in exoplanetary systems. 

The structure of this paper is as follows: Section \ref{Target Selection WASP-12b} describes the target selection. Section \ref{Observation log} presents the observational data collected from various sources, including both space-based and ground-based observations, along with details of data processing and published light curves from the literature. Section \ref{Modeling WASP-12b} outlines the methods and procedures employed to analyze all 391 light curves. Section \ref{Transit timing analysis WASP-12b} provides an overview of the transit timing analysis, in which three timing models are fitted to the data. In Section \ref{Discussions}, we discuss the results of the transit timing analysis. Finally, Section \ref{Conclusions} summarizes the conclusions of this work.

\section{Target Selection}
\label{Target Selection WASP-12b}

WASP-12\,b continues to be among the most thoroughly investigated exoplanets to date. \citet{2011A&A...528A..65M} initially reported short-term transit timing deviations in WASP-12\,b, attributing them to dynamical perturbations from a potential unseen planetary companion. However, subsequent studies did not confirm these short-term variations; instead, they revealed evidence of long-term timing deviations in the system. \citet{2016A&A...588L...6M} provided the first confirmation of timing offsets in WASP-12\,b, identifying a declining trend consistent with orbital decay. This result was further supported by independent analyses incorporating additional data (\citealt{2017AJ....154....4P, 2017AJ....153...78C, 2018AcA....68..371M, 2019MNRAS.490.1294B}). While the long-term trend could also be interpreted through alternative mechanisms such as the R$\phi$mer effect or apsidal precession, \citet{2019MNRAS.482.1872B} suggested apsidal precession as a plausible explanation. In contrast, \citet{2020ApJ...888L...5Y} presented strong evidence ruling out both apsidal precession and the R$\phi$mer effect as dominant contributors, thereby reinforcing the orbital decay hypothesis.

Previous theoretical studies by \citet{2017MNRAS.470.2054C} and \citet{2017ApJ...849L..11W} suggested that the observed orbital decay of WASP-12\,b could be driven by dissipation of tidally excited internal gravity waves (IGWs), contingent on the adopted stellar structure. In these models, efficient dissipation arises through wave breaking in a fully radiative stellar core. However, this mechanism requires WASP-12 to be a subgiant star, whereas observational evidence indicates that it is a main-sequence F-type star with a convective core, for which IGW wave breaking is not expected to operate. Subsequent work by \citet{2020MNRAS.498.2270B}, as well as a broader parameter survey by \citet{10.1093/mnras/sty2805}, found no stellar model capable of simultaneously satisfying all observational constraints, highlighting a tension between tidal theory and observations.  

More recently, analyses incorporating both primary and extended \textit{TESS} observations \citep{2021AJ....161...72T, 2022AJ....163..175W} have strengthened the empirical evidence for orbital decay in WASP-12\,b. This apparent discrepancy between theory and observations may be alleviated by a newly proposed tidal dissipation mechanism involving the conversion of IGWs into magnetic waves through interaction with a magnetic field generated by a convective core dynamo \citep[see also][]{2024ApJ...966L..14D}. In this framework, efficient tidal dissipation can occur in F-type stars with convective cores over certain evolutionary stages, particularly at older ages. Given these ongoing theoretical developments and the availability of more than a decade of high-precision transit timing data, WASP-12\,b remains a prime laboratory for detailed transit timing studies.

\section{Observation Log}
\label{Observation log}
For WASP-12\,b, we analyzed photometric transit data from both space and ground based observatories to enhance transit timing precision and extend the temporal coverage. We obtained space based observations from the Transiting Exoplanet Survey Satellite (TESS; \citealt{2014SPIE.9143E..20R}) and re-processed the data. As for ground based observations, in addition to seven new photometric observations, collected using the 0.61 m VASISTHA Telescope, the 1.3 m Devasthal Fast Optical Telescope (DFOT), and the 0.3 m AG Optical IDK telescope, we have also sourced some ground based light curves from public databases, the Exoplanet Transit Database (ETD; \citealt{2010NewA...15..297P}) and the ExoClock Project (\citealt{2023ApJS..265....4K}). We obtained 97 high-quality (DQ 1 and 2) light curves from ETD and 34 light curves from the ExoClock Project. Furthermore, 134 additional published light curves were incorporated, with details provided in the following sections. Together, these datasets enable a robust transit timing analysis by combining the high precision of space-based observations with the extended temporal baseline afforded by ground-based monitoring.

\subsection{TESS observations}

The Transiting Exoplanet Survey Satellite (TESS) was launched on 18 April, 2018 aboard a SpaceX Falcon 9 to search for exoplanets using the transit method. Designed as the successor to Kepler and K2, TESS surveys an area nearly 400 times larger than Kepler's field of view, monitoring about 200,000 bright nearby stars.

Equipped with four wide-field cameras providing a $24^{\circ} \times 96^{\circ}$ field of view, TESS observes each sector for about 27 days, enabling an almost all-sky survey. During its prime two-year mission (2018–2020), TESS covered $75\%$ of the sky, discovering 66 confirmed exoplanets and identifying over 2100 candidates. These include hot Jupiters, sub-Neptunes, and Earth-sized rocky planets, some within the habitable zone. Notable finds include TOI-700~d (Earth-sized with habitable zone, \citealt{2020AJ....160..116G}), Pi~Mensae~c (super-Earth, \citealt{2018A&A...619L..10G}), and LHS~3844~b (Earth-sized, a key target for JWST studies, \citealt{2019ApJ...871L..24V}). 

Although its prime mission ended in July 2020, TESS continues in an extended mission, with over 400 confirmed exoplanets and thousands of candidates (as of 2025). Its high-precision transit photometry supports atmospheric studies, orbital dynamics, and JWST target selection, making TESS a cornerstone of modern exoplanet science and a key step toward identifying potentially habitable worlds (\citealt{2014SPIE.9143E..20R}).

TESS first observed WASP-12 (TIC 86396382) during Sector~20 of its primary mission, covering the period from 2019 December 24 to 2020 January 21 (UT). In this study, we analyzed a total of 391 mid-transit times, including 119 full transit light curves extracted from TESS. The target was observed using Camera~1 with a two-minute cadence across six sectors, 20, 43, 44, 45, 71, and 72, over the time interval from 2019 February 28 to 2023 March 10, with exposure times ranging from 2 to 30 minutes. Observations were performed using the TESS photometer mounted on the telescope. Table \ref{tab:1 sector} presents the count of complete transits recorded and the associated data points retrieved from the TESS database for each sector. We have presented the number of full transits observed and the corresponding data points obtained from the TESS database for each sector in Table \ref{tab:1 sector}.

The data were downlinked from the spacecraft and processed through the Science Processing Operations Center (SPOC) pipeline at NASA Ames Research Center (\citealt{2016SPIE.9913E..3EJ}). The SPOC pipeline determined the optimal photometric aperture, extracted the light curves, and applied systematic corrections to produce science-ready data products suitable for transit analysis. All final data products are publicly available through the Mikulski Archive for Space Telescopes (MAST).

\begin{table*} [htbp]
\begin{center}
\caption {Log of Sector specific TESS photometric observations of WASP-12 utilized in our analysis} 
\label{tab:1 sector}
\begin{tabular}{cccc}
\hline
Object Name & TESS sector & Number of Transits & Data points\\
\hline
 & No. 20   & 20 &  16552\\
{WASP-12} & No. 43   & 27		&	15577\\
& No. 44  & 20		&	15779\\
& No. 45   & 21		&	16085\\
& No. 71   & 20		&	16627\\
& No. 72   & 19		&	15059\\
\hline
\hline
{Total} &  & 119 & 95679\\
\hline
\hline
\end{tabular}\\
\end{center}
\end{table*}

\subsubsection{Data Reduction of high-precision TESS data}
We retrieved TESS light curves from the NASA funded astronomical data archive repository Mikulski Archive for Space Telescopes (MAST)\footnote{All the TESS data used in this paper can be found in MAST \citep{10.17909/t9-nmc8-f686}.}, using the Presearch Data Conditioning Simple Aperture Photometry (PDCSAP) products generated by the Science Processing Operations Center (SPOC; \citealt{2016SPIE.9913E..3EJ}). Compared to standard SAP data, PDCSAP light curves reduce scatter, suppress short-term noise, and correct long-term instrumental trends (\citealt{2012PASP..124.1000S, 2012PASP..124..985S, 2014PASP..126..100S, 2020RNAAS...4..201C}).

Transit windows were extracted within $\pm 0.125$~days of the predicted mid-transit times. Incomplete or noisy transits were excluded, as missing ingress or egress hampers precise timing (\citealt{2013MNRAS.430.3032B}). Data pre-processing, including extraction, normalization, and detrending, was performed with the \texttt{JULIET} package (\citealt{2019MNRAS.490.2262E}), which employs Bayesian inference through the \texttt{MultiNest} algorithm (\citealt{2009MNRAS.398.1601F, 2019OJAp....2E..10F, 2016S&C....26..383B}). We retained only points with a quality flag of zero and converted time stamps to $\mathrm{BJD_{TDB}}$ by adding 2,457,000 (\citealt{2010PASP..122..935E}).

Stellar and instrumental variability was modeled using a Gaussian Process (GP) with a Matérn kernel implemented in \texttt{CELERITE} (\citealt{2019MNRAS.490.2262E}). A normal prior was set for the mean out-of-transit flux, the dilution factor fixed to unity, and wide log-uniform priors adopted for GP hyperparameters (amplitude, timescale, and jitter). This procedure yielded GP-corrected light curves, optimized for precise transit timing and exoplanet characterization. Other details are given in Section \ref{Modeling WASP-12b}.

\subsection{New Ground-Based Light Curves}
For WASP-12 b, we obtained one transit on 7 January, 2025 using the 1.3 m Devasthal Fast Optical Telescope (DFOT) at the Devasthal, Nainital campus of ARIES. In addition, three transits were observed on 26 and 27 December, 2024 and 24 February, 2025 with the 0.61 m reflecting telescope, VASISTHA at the IERCOO campus of ICSP, Kolkata. Three further transits were recorded on 2 January, 4 January, and 26 March 2025 with the 0.3 m AG Optical IDK telescope at the Utah Desert Remote Observatory (UDRO) in Beryl, Utah.  The exposure times were optimized to compensate for variable weather conditions. The data reduction procedures have been described in detail in Section~\ref{Data reduction of ground based data}.  

\subsubsection{Devasthal Fast Optical Telescope}
The observations of one new transit of WASP-12 b were conducted using the 1.3-m Devasthal Fast Optical Telescope (DFOT), located at the Aryabhatta Research Institute of Observational Sciences (ARIES), Nainital, India. The DFOT, a modern Ritchey--Chrétien Cassegrain telescope with a 1.3 m aperture, was installed at Devasthal by DFM Engineering Inc., USA, and is operated by ARIES, an autonomous institute under the Department of Science and Technology (DST), Government of India. Equipped with a fast $f/4$ optical system, the telescope is mounted on a fork--equatorial system with single-axis tracking. Its secondary mirror is controlled by a five-axis actuator for precise focusing, while friction drives provide backlash-free motion in right ascension and declination, ensuring a pointing accuracy of $\sim 10$~arcsec (rms). For imaging, a back-illuminated CCD camera with $2048 \times 2048$ pixels ($13.5\,\mu\mathrm{m}$ pixel size) and deep thermoelectric cooling down to $-80^{\circ}$C was used. The transit observation using DFOT was carried out using an $R$-band filter.

\subsubsection{Ionospheric and Earthquake Research Centre and Optical Observatory}
We observed three transits of the hot Jupiter WASP-12\,b with the 0.61 m VASISTHA reflecting telescope at the Ionospheric and Earthquake Research Centre and Optical Observatory (IERCOO), established by ICSP at Sitapur, Paschim Medinipur, West Bengal. This telescope, the largest in eastern India, is mounted on an Ascension 200 German equatorial system with high-resolution encoders. All observations were conducted in the Cousins $R$ band to reduce limb darkening, color-dependent extinction, and to allow high-cadence photometry (\citealt{2007ApJ...664.1185H}). The telescope employs a 24 inch primary mirror with a focal ratio of $f/6.5$ and is equipped with an Atik 460EX Mono CCD camera. The detector has $2749 \times 2199$ pixels of size $4.54~\mu\mathrm{m}$, providing an image scale of $0.235~\mathrm{arcsec~pixel}^{-1}$.

\subsubsection{Utah Desert Remote Observatory}
We conducted observations of three new transits of the exoplanet WASP-12\,b using the 0.3 m AG Optical imaging Dahl Kirkham (IDK) telescope at the Utah Desert Remote Observatory (UDRO) in Beryl, Utah. The telescope is equipped with a 12.5 inch primary mirror with a diameter of 318~mm and a focal ratio of $f/6.7$, corresponding to a focal length of 2128~mm. It is paired with a ZWO ASI294 MM CCD camera. The detector has $4144 \times 2822$ pixels of size $4.63~\mu\mathrm{m}$, providing an image scale of $0.451~\mathrm{arcsec~pixel}^{-1}$. All observations were performed using $R$, and $I$ band filters.

The complete log of all new photometric observations, including the filters employed and the corresponding exposure times, is presented in Table~\ref{tab:1 WASP-12b}. And the specifications of the above mentioned telescopes and CCD detectors are mentioned in Table~\ref{tab: Specifications of Telescopes WASP-12 b}.

\begin{table}[htbp]
\centering
\caption{Details of the New Photometric Observations for  WASP-12b}
\label{tab:1 WASP-12b}
\begin{tabular}{llcccl}
\hline\hline
Date of Obs & Telescope & \(N_{\text{frames}}\) & Exp(s) & Filter & Epoch \\
\hline
2025 Jan 7 & 1.3-m & 418 & 20 s & R & 5652\\
2024 Dec 26 & 0.61-m & 283 & 40 s & R & 5640\\
2024 Dec 27 & 0.61-m & 253 & 40 s & R & 5641\\
2025 Feb 24 & 0.61-m & 783 & 15 s & R & 5695\\
2024 Jan 2 & 0.3-m & 199 & 90 s & R & 5647\\
2024 Jan 4 & 0.3-m & 162 & 90 s & R & 5648\\
2025 Mar 26 & 0.3-m & 84 & 180 s & I & 5723\\

\hline
\end{tabular}
\end{table}

\begin{table}[htbp]
\centering
\caption{Specification of Telescopes and CCD Detectors Used in This Work}
\label{tab: Specifications of Telescopes WASP-12 b}
\begin{tabular}{cccccc}
\hline\hline
Telescope and CCD Detector & CCD Size & Field of View  & Plate Scale & Readout Noise  & Gain \\
& (pixels) & ($ \text{arcmin} \times \text{arcmin} $) & ($\text{arcsec pixel}^{-1}$) & ($e^{-1}$) & ($\mathrm{e}^{-1}/\mathrm{ADU}$)\\
\hline
1.3 m DFOT, Andor's DZ436 CCD  & $2048 \times 2048$ & $18 \times 18$  & 0.535 & 7.0 & 2.0\\
0.61 m VASISTHA, ATIK 460EX Mono CCD  & $2749 \times 2199$ & $10.8 \times 8.6$  & 0.235 & 5 & 0.27\\
0.3 m IDK, ZWO ASI294 MM pro  & $4144 \times 2822$ & $31 \times 21$  & 0.451 & 1.8 & 1.0\\
\hline
\end{tabular}
\end{table}

\subsubsection{Data Reduction of New Ground-based Data}
\label{Data reduction of ground based data}

The telescopes record raw CCD frames, and atmospheric turbulence, optical distortions, focus errors, charge diffusion, and detector electronics affect them. We corrected these frames through image processing to extract accurate stellar positions and brightness for exoplanet transit studies. This process involves three phases: pre-processing, processing, and post-processing.

\paragraph{Pre--processing:}

Before performing aperture photometry, we preprocessed the raw CCD images to remove instrumental and observational artifacts. We corrected defects such as bad pixels, cosmic ray hits, and pixel to pixel sensitivity variations to ensure reliable photometric measurements. Along with the science frames, calibration frames namely bias, dark, and flat fields are obtained to remove instrumental effects from the CCD images, including electronic offsets and thermal current, and to correct for pixel-to-pixel sensitivity variations and optical effects. Using IRAF\footnote{The Image Reduction and Analysis Facility (IRAF) is a software package distributed by the National Optical Astronomy Observatory (NOAO), which is operated by the Association of Universities for Research in Astronomy (AURA) under a cooperative agreement with the National Science Foundation (NSF).} (Image Reduction and Analysis Facility), we apply standard routines such as zerocombine, flatcombine, and ccdproc under the imred package of NOAO to perform trimming, bias subtraction, dark correction, and flat--fielding, and we remove cosmic rays with the cosmic rays task in crutil. After calibration, we inspected the images for residual defects and alignment issues, displayed them with the task display, and interactively analyzed them using imexamine to estimate parameters such as the full width at half maximum (FWHM) of stellar point spread functions and the local sky background. These parameters are essential for defining the photometric aperture radius and sky annulus, which directly affect the precision of the extracted light curves in exoplanet transit analysis.

\paragraph{Processing}: We performed the aperture photometry using the \texttt{apphot} and \texttt{daophot} packages in IRAF. First, we defined the photometry parameters in \texttt{datapars}, \texttt{centerpars}, \texttt{fitskypars}, and \texttt{photpars} to specify CCD characteristics, centering methods, sky background estimation, and aperture sizes. We prepared a coordinate file for the target star and nearby comparison stars, although stars could also be selected interactively. Aperture photometry was then applied across all science frames using the \texttt{phot} task to extract instrumental magnitudes, with magnitudes computed for the chosen aperture(s). The aperture radii were typically set to 2-3 times the full width at half maximum (FWHM) of the stellar point-spread function (PSF) and optimized to minimize scatter in the out-of-transit (OOT) light curve. We tracked stellar centroids across all frames, typically by cross-correlating with a reference image, to ensure precise aperture alignment during extraction. Photometry was then performed with the \texttt{daophot} package in IRAF on the target star and 2-8 nearby comparison stars of similar brightness and color. The aperture configuration and number of comparison stars were adjusted slightly from night to night according to atmospheric conditions and field orientation, following the approach of \citet{2016AJ....151...17J}.

\paragraph{Post--processing:}

Once we extracted the instrumental magnitudes, we compiled the results using \texttt{txdump} and related tools to obtain magnitudes, errors, and centroids for the target and comparison stars. For exoplanet transit studies, we performed differential photometry by constructing a reference flux from the summed fluxes of the chosen comparison stars and dividing the target star's flux by this reference. This produced a time series of relative flux, forming the transit light curve. Then, we converted all time stamps to the Barycentric Julian Date in the Barycentric Dynamical Time system (BJD\(_{\mathrm{TDB}}\)) using the publicly available code of \citet{2010PASP..122..935E}. Further modeling and fitting, allowed us to refine the stellar and planetary parameters of the exoplanet, which are described in detail in Section \ref{Modeling WASP-12b}.

The original data points across all seven nights are presented in Table \ref{tab:longtable_captionlable}.

\begin{center}
\small\addtolength{\tabcolsep}{4.0pt}
\begin{longtable*}{cccccc}
\caption{Newly Observed Transit Light-Curve Data of WASP-12 b Over Seven Nights} \label{tab:longtable_captionlable}\\
\hline
\hline
Target Name & Telescope & Epoch & TDB-based BJD & Normalized Flux & Normalized Flux Error\\
\hline
\endfirsthead
WASP-12 b  & 1.3 m & 5652 & 2460684.12708776 & 1.0015969 & 0.0031616\\
WASP-12 b  & 1.3 m & 5652 & 2460684.12749167 & 1.005571 & 0.0031616\\
WASP-12 b  & 1.3 m & 5652 & 2460684.12789558 & 1.0047181 & 0.0031616\\
... & ... & ... & ... & ...\\
\hline
\end{longtable*}
Note. This table is available in its entirety in machine-readable form. A portion is shown here for guidance regarding its form and content.
\end{center}

\subsection{Ground-based Observations from Public Databases}
\subsubsection{Exoplanet Transit Database}
The ETD, established in September 2008, serves as a collaborative platform where amateur astronomers worldwide contribute transit measurements. As of January 4, 2026, the database contains 84,307 observations contributed by 1737 observers from different observatories across the globe. In addition to the TESS observations, we incorporated complete transit light curves from the Exoplanet Transit Database (ETD), selecting only those with a data quality index ($DQ < 3$) to ensure consistency and reliability.  These high-quality, community sourced datasets provided an essential complement to the TESS data, thereby improving the robustness of our transit timing analysis. Currently, all original ETD light curves are available on the VarAstro\footnote{https://var.astro.cz/en/} server, which was launched last year. 

\subsubsection{ExoClock Project}
The ExoClock project, initiated in September 2019, is an interactive platform designed to coordinate the regular monitoring of transiting exoplanets with small to medium-sized telescopes. Its primary objective is to refine the ephemerides of targets for ESA's upcoming Ariel mission, while also fostering international collaboration to improve the temporal coverage and precision of transit measurements.

In this study, we incorporated only complete light curves from ExoClock, which were analyzed to determine their mid-transit times. These were then combined with transit data from ETD to perform a comprehensive transit timing analysis. Although the number of ExoClock observations is relatively limited, they provide crucial supplementary coverage, thereby extending the time baseline. Future work will integrate additional data from both ETD and ExoClock to further enhance the accuracy of orbital parameter refinements.

\subsection{Published Ground-Based Light curves}
As described above, in addition to the 7 new transit light curves and 250 light curves obtained from TESS, ETD, and the ExoClock Project, we have compiled and analyzed a total of 108 publicly available complete light curves. These include 3 from \citet{2024PSJ.....5..163A}, 2 from \citet{2011AJ....141..179C}, 21 from \citet{Leonardi_2024}, 2 from \citet{Maciejewski_2011}, 44 from \citet{Maciejewski_2013}, 28 from \citet{Maciejewski_2016}, and 8 from \citet{2020ApJ...888L...5Y}. Additionally, we obtained light curves from the authors Hebb Lesbie, Efrain Alavarado, and Karen Collins via private communication and applied selection criteria to include only those that are complete, have an out-of-transit baseline, and possess timing uncertainties of less than 5 minutes. Finally, we incorporated 23 light curves from \citet{2017AJ....153...78C}, 2 from \citet{2009ApJ...693.1920H}, and 1 from \citet{2024MNRAS.534..800A}, adding these 26 light curves to the existing database for analysis. The complete list of all 391 light curves is provided in Table~\ref{tab:2 WASP-12b}.

\begin{table}
\begin{center}
\caption{Details of all 391 transit light curves of WASP-12 b considered in this work.}
\label{tab:2 WASP-12b}
\begin{tabular}{cc}
\hline
Number of light  & Sources  \\
curves taken &  \\
\hline
  119 & TESS\\
 97 & ETD \\
34 & Exoclock \\
 44 & \citet{Maciejewski_2013} \\
 28 & \citet{Maciejewski_2016} \\
 23 & \citet{2017AJ....153...78C} \\
21 & \citet{Leonardi_2024} \\
 3 & IERCOO \\
 3 & UDRO \\
 ~3 & \citet{2024PSJ.....5..163A} \\
 ~2 & \citet{2011AJ....141..179C} \\
 ~2 & \citet{2009ApJ...693.1920H} \\
 ~2 & \citet{Maciejewski_2011} \\
 ~8 & \citet{2020ApJ...888L...5Y} \\
 ~1 & DFOT \\
 ~1 & \citet{2024MNRAS.534..800A} \\

\hline
\end{tabular}
\end{center}
\end{table}

\section{Transit Light-curve Fitting}
\label{Modeling WASP-12b}
To perform a precise and homogeneous transit-timing analysis, and to determine the individual mid-transit times ($T_{\mathrm{m}}$) while refining the stellar and planetary parameters of the WASP-12 system, we modeled all 391 transit light curves using \texttt{juliet} \citep{2019MNRAS.490.2262E}. The transit light curves were modeled simultaneously with a Gaussian Process (GP) noise model. To do this, we adopted a simple, approximate Matérn kernel, implemented using the \texttt{celerite} formalism. The covariance between the $i$-th and $j$-th data points is described by

\begin{equation}
k(\tau_{i,j}) = \sigma_{GP}^{2}\,\tilde{M}(\tau_{i,j}, \rho) 
+ (\sigma_i^{2} + \sigma_w^{2})\,\delta_{i,j},
\end{equation}

where $k(\tau_{i,j})$ represents the $(i,j)$ element of the covariance matrix $\Sigma$, and 
$\tau_{i,j} = |t_i - t_j|$ is the temporal separation between the GP regressors $t_i$ and $t_j$, 
which in this work correspond to the observation times. The term $\sigma_i$ denotes the 
measurement uncertainty of the $i$-th data point, $\sigma_{GP}$ sets the GP amplitude (in ppm), 
and $\sigma_w$ (in ppm) represents an additional, unknown white-noise \textit{jitter} component. 
The symbol $\delta_{i,j}$ denotes the Kronecker delta, which is equal to unity when $i=j$ and zero 
otherwise. The function $\tilde{M}(\tau_{i,j}, \rho)$ is given by

\begin{equation}
\tilde{M}(\tau_{i,j}, \rho) =
\left[
(1 + 1/\epsilon)\exp\!\left(-[1 - \epsilon]\sqrt{3}\tau/\rho\right)
+
(1 - 1/\epsilon)\exp\!\left(-[1 + \epsilon]\sqrt{3}\tau/\rho\right)
\right],
\end{equation}

which corresponds to the approximate Matérn component of the kernel and is characterised by a correlation length-scale $\rho$. This approach is particularly advantageous as it accounts for correlated (non-white) noise present during transit, thereby yielding more realistic uncertainties on the inferred planetary parameters. Within \texttt{juliet}, we simultaneously fitted the transit parameters and the GP hyperparameters, adopting appropriate priors and providing the relevant GP regressors. To correct for atmospheric transparency variations, we normalized the light curves by fitting a linear function to the out-of-transit (OOT) data during the fitting stage and including this OOT model simultaneously with the transit model. This approach avoids artificially removing degrees of freedom from the light-curve data and underestimating the uncertainties of the final parameters.

In addition to our newly obtained transit observations and the TESS, ETD, and ExoClock light curves, we used archival light curves obtained either from the published literature or through private communication with the respective authors.

To ensure the homogeneity and reliability of the dataset, we applied the following selection criteria before incorporating any unpublished ground-based light curves, as well as published light curves from the literature, ETD, and ExoClock, into our analysis:
\begin{enumerate}
    \item Only complete and high-quality transit light curves were included.
    \item Light curves with timing uncertainties greater than 5 minutes were excluded.
\end{enumerate}

Each WASP-12\,b transit light curve was analyzed individually using \texttt{juliet} in order to determine the corresponding mid-transit time. In our fits, we sampled the transformed parameters $r1\_p1$ and $r2\_p1$, which parameterize the planet-to-star radius ratio, $p = R_{\rm p}/R_\ast$, and the impact parameter, defined as $b = (a/R_\ast)\cos i$. We also fitted the limb-darkening parametrization $q_{1,\mathrm{TESS}}$ and $q_{2,\mathrm{TESS}}$, which parameterize the quadratic limb-darkening coefficients $u_1$ and $u_2$ following the quadratic law of \citet{2013MNRAS.435.2152K}. The transformation from the $(r_1, r_2)$ plane to the $(b, p)$ plane was implemented using the formalism described by \citet{2018RNAAS...2..209E}.

For the \textit{TESS} light curves, the initial values of quadratic limb-darkening coefficients ($u_1$, $u_2$) were obtained by interpolating the stellar effective temperature ($T_{\mathrm{eff}}$), surface gravity ($\log g$), metallicity ($[\mathrm{Fe/H}]$), and microturbulent velocity ($V_{\mathrm{t}}$) from the limb-darkening tables of \citet{2017A&A...600A..30C}. For all other ground-based transit light curves observed in the clear, $V$, $I$, luminance, and $R$ filters, as well as the Sloan $r$ and $z$ bands, we followed the procedure of \citet{2021AJ....161..108S} and interpolated the quaratic limb-darkening coefficients from the tables of \citet{2011A&A...529A..75C} using the \texttt{EXOFAST} package (\citealt{2013PASP..125...83E}).

The adopted stellar parameters were $T_{\mathrm{eff}} = 6265\,\mathrm{K}$, $\log g = 4.11\,\mathrm{cm\,s^{-2}}$, and $[\mathrm{Fe/H}] = 0.12$, as reported by \citet{2024A&A...686A..84L}. Since the clear filter spans both the $V$ and $R$ passbands (\citealt{Maciejewski_2013}, the limb-darkening coefficients for light curves obtained in this filter were taken as the average of the corresponding $V$ and $R$ band values. For light curves observed with the luminance filter, the coefficients derived for the $V$ band were adopted.

For the dataset of \citet{Maciejewski_2013}, the limb-darkening coefficients ($u_1$, $u_2$) reported in their study were used directly. For the 23 light curves from \citet{2017AJ....153...78C}, we followed their methodology and adopted coefficients calculated for the $R$ and \textit{Kepler} passbands to represent their Clear with Blue Block (CBB; a high-pass filter with a cutoff near 500~nm) and open filters, respectively. For \citet{Maciejewski_2016}, the coefficients provided for the Clear filter were adopted as representative for the None filter.

The prior distributions adopted for the planetary and instrumental parameters are listed in Table~\ref{tab:3 WASP-12b}, while the parameterized limb-darkening coefficients ($q_1$ and $q_2$) for the different filters are summarized in Table~\ref{tab:4 WASP-12b}.

\begin{table} [htbp]
\begin{center}
\caption{The Initial Values set for the Parameters}
\label{tab:3 WASP-12b}
\begin{tabular}{ccc}
\hline
\hline
Name of Parameters & Description & Prior \\
\hline
\hline
Planetary Parameters & &\\
P (days) & Orbital period & $\mathcal{N}(1.0914210,\;0.00000020)^{a,\dagger}$ \\
${T_m}$ (days) & Time of transit-centre & $\mathcal{U}($tmid1$,$tmid2$)^{\S, \dagger}$\\
r1\_p1$^{\ast}$ & Transformed transit parameter (jointly parameterizes $b$ and $R_{\mathrm{p}}/R_\star$) & $\mathcal{U}(0.37215959921,0.39005188)$  \\
r2\_p1$^{\ast}$ &Transformed transit parameter (jointly parameterizes $b$ and $R_{\mathrm{p}}/R_\star$) & $\mathcal{U}(0,1)$\\
e & Eccentricity of the orbit & $\mathrm{Fixed}\,(0.0)$\\
$\omega$ (degree) & Argument of periastron & $\mathrm{Fixed}\,(90\degree.0)$\\
\hline
Instrumental parameters &&\\
$\rho$ & Mean stellar density & $\mathcal{N}(0,\,0.1)$ \\
mdilution & Dilution factor & $\mathrm{Fixed}\,(1.0)$ \\
mflux & Relative flux offset & $\mathcal{N}(0,\,0.1)$ \\
sigma\_w & jitter & $\mathcal{N}(0,\,0.1)$ \\
$\sigma_{GP}$ & Amplitude of the GP & $\log\mathcal{U}(10^{-6},\,10^{6})^{\dagger}$\\
$\rho_{\mathrm{GP}}$ & Time/length-scale of the Matern part of the GP & $\log\mathcal{U}(10^{-3},\,10^{3})$\\
$q_1$ & Parameterized limb-darkening coefficient &   $\mathcal{N}(${According to filter}$,\;0.05)^b$\\
$q_2$ & Parameterized limb-darkening coefficient &   $\mathcal{N}(${According to filter}$,\;0.05)^b$\\
\hline
\end{tabular}
\end{center}
{Notes:
$^a$ The initial and the prior value of the parameter P is directly adopted from \citet{Leonardi_2024}.\\
$^b$ The priors of $q_1$ and $q_2$ taken from \citet{2017A&A...600A..30C}.\\
$^\dagger$ $\mathcal{N}(\mu,\sigma)$ represents a normal prior, $\mathcal{U}(a,b)$ a uniform prior, and $\log\mathcal{U}(a,b)$ a log-uniform prior.\\
$^\S$ Here, tmid1 and tmid2 are the lower and upper bounds for mid-transit time for each light curve.\\
* r1\_p1 and r2\_p1 are the reparameterized transit parameters following \citet{2019MNRAS.490.2262E}. The quadratic limb-darkening coefficients are parameterized as $(q_1, q_2)$ following \citet{2013MNRAS.435.2152K}.
}\\
\end{table}

\begin{table*} [htbp]
\begin{center}
\caption {The Calculated Values for parameterized Limb-darkening Coefficients}
\label{tab:4 WASP-12b}
\begin{tabular}{lcc}
\hline
Filter & $q_1$ & $q_2$\\
\hline
{\it V} & 0.4690 & 0.2783\\
 {\it R} & 0.3741 & 0.2395\\
 {\it I} & 0.2808 & 0.2072\\
 Clear & 0.4202  & 0.2600\\
 Sloan r & 0.4016  &  0.2481\\
 Sloan z & 0.2393  & 0.1906\\
 Luminance & 0.4690 & 0.2783\\
 TESS & 0.4560 & 0.3293\\
\hline

\end{tabular}\\
\label{tb:LD_coefficients}
\end{center}
\end{table*}

\begin{figure}
    \centering
    \includegraphics[width=1.1\linewidth]{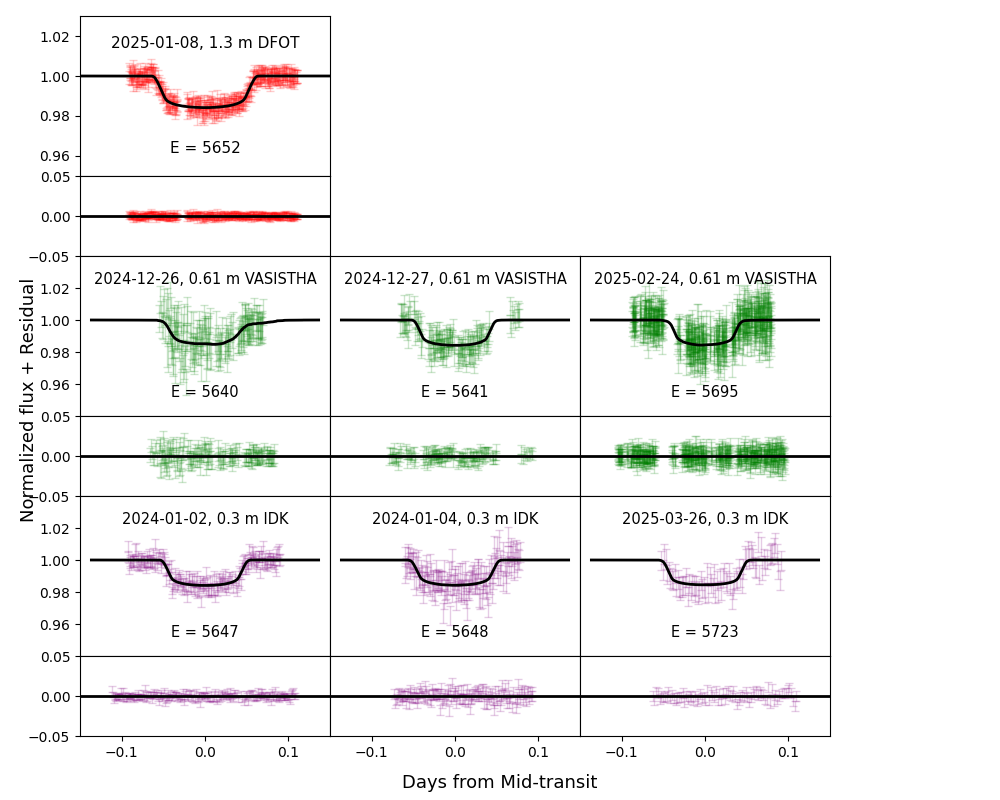}
    \caption{Seven new light curves of WASP-12\,b observed with the 1.3\,m Devasthal Fast Optical Telescope (top), 0.61\,m VASISTHA (middle), and the 0.3\,m IDK telescope (bottom). The upper panels show the normalized light curves, while the lower panels display the corresponding residuals. The data shown have been corrected by subtracting the best-fitting Gaussian Process (GP) model and the out-of-transit (OOT) linear baseline function. The solid line represents the best-fitting transit model obtained through juliet. Flux uncertainties are indicated by error bars, shown in red, green, and purple for the three telescopes, respectively.}
    \label{fig:all in one 7 new LC}
\end{figure}

For each fitted parameter, the 50th percentile of the posterior probability distribution was adopted as the best-fit value, while the 15.9th and 84.1st percentiles were taken as the lower and upper $1\sigma$ uncertainties, corresponding to the 68\% credible intervals. Graphical representation of the newly acquired 7 normalized light curves from various ground-based observatories is shown in Figure \ref{fig:all in one 7 new LC}. While those from \textit{TESS}, ETD, and ExoClock are presented in Figures \ref{fig:TESS1_WASP-12b}-\ref{fig:TESS5_WASP-12b}, \ref{fig:ETD1_WASP-12b}-\ref{fig:ETD2_WASP-12b}, and \ref{fig:Exoclock1_WASP-12b}-\ref{fig:Exoclock2_WASP-12b}, respectively.

\section{Timing Analysis}
\label{Transit timing analysis WASP-12b}

Before initiating the transit-timing analysis, we identified seven transits that are reported in both
the ETD and ExoClock databases. They are obviously the same light curves as the observer names, telescopes, and observatory information are identical, in addition, the observation dates and filters are also the same. Observers submitted the same light curves to both ETD  and ExoClock databases. We verified that the corresponding mid-transit times from the two databases are consistent within their quoted uncertainties. To avoid double-counting the same observational epochs, we retained a single representative mid-transit time per transit in the final timing analysis.

In these cases, we adopted the mid-transit times derived from our fits to the ETD light curves as the preferred values, since the corresponding light curves exhibit superior quality, characterized by lower noise levels, smaller timing uncertainties, and Data Quality (DQ) indices below 3.

After applying the above mentioned correction, we modeled the transit timing data using three distinct approaches, following the methodologies described in \citet{2017AJ....154....4P} and \citet{2020ApJ...888L...5Y}. In this analysis, we combined all 391 mid-transit times obtained from both space-based and ground-based observations.The first model is the standard constant-period assumption, represented by a linear ephemeris. The linear ephemeris is expressed as,  

\begin{equation}
    T_{c}(E) = T_{0} + E P ,
\end{equation}

where $E$ is the transit epoch, $T_{0}$ is the reference mid-transit time at epoch $E = 0$ (taken as the first transit of WASP-12 b observed by \citealt{2009ApJ...693.1920H}), $P$ is the orbital period, and $T_{c}(E)$ is the calculated mid-transit time at epoch $E$.

To refine the best-fit parameters obtained from linear fitting, we adopted a Gaussian likelihood with uniform priors on $P$ and $T_0$ and sampled the posterior using the affine-invariant MCMC algorithm of \citet{2010CAMCS...5...65G}. Convergence and sampling efficiency were assessed using the mean acceptance fraction ($a_f$), integrated autocorrelation time ($\tau$), and effective sample size ($N_{\rm eff}$). The chains exhibit robust convergence, with $a_f \simeq 0.44$ (within the optimal range 0.2--0.5) and $N_{\rm eff} > 50$ per walker, satisfying standard MCMC criteria \citep{2013PASP..125..306F, 2017ApJ...848....9S}. MCMC non-convergence can lead to unstable model behavior \citep{2022AJ....164..220H}. We therefore evaluated convergence using the integrated autocorrelation time ($\tau$) \citep{2010CAMCS...5...65G, 2013PASP..125..306F}, obtaining $\tau \simeq 19$ steps. The effective number of independent samples is $N_{\rm eff} \simeq 1053$, exceeding the recommended minimum threshold of 50 per walker, as advised by the \texttt{emcee} framework. Then we calculated the timing residuals, O-C (pronounced ``O minus C''; \citealt{2005ASPC..335....3S}), where O denotes the observed mid-transit times and C is the calculated mid-transit times. In the absence of transit timing variation, we would anticipate no significant deviations of the derived O-C (observed-minus-calculated) values from zero. However, we observed a significant deviation on both sides of zero and this deviation from the zero value could be the ﬁrst indication of timing anomalies. The estimated timing residuals (O-C) along with their corresponding epochs and original mid-transit times ($T_m$) are shown in Table \ref{tab:6}.

\begin{center}
\small\addtolength{\tabcolsep}{-3pt}
\begin{longtable*}{cccccc}
\caption{Mid-transit Times $({T}_{m})$ and Timing Residuals (O-C) for all 391 Transit Light Curves of WASP-12 b} 
\label{tab:6}\\
\hline
\hline 
Transit Number & \ \ \ \ ${T}_{m}$  & $O-C$ & Light Curve VarAstro ID & \ Transit Source & Timing Source\\
({E}) & \ \ \ \ (BJD$_{\rm TDB}$) & \ \ (days)  & & & \\
\hline
\endfirsthead
0 & 2454515.52548 & -0.0033471 & ... & \citet{2009ApJ...693.1920H} & This Paper \\
13 & 2454529.71660 & -0.0006704 & ... & \citet{2009ApJ...693.1920H} & This Paper \\
 ...&...&...&...&...&...\\
\hline
\end{longtable*}
Note. This table is available in its entirety in machine-readable form.\\
References. \citet{2009ApJ...693.1920H, 2011AJ....141..179C, 2011A&A...528A..65M, 2013A&A...551A.108M, 2016A&A...588L...6M, 2017AJ....153...78C, 2020ApJ...888L...5Y, 2024PSJ.....5..163A, 2024MNRAS.534..800A}.
\end{center}

The second model corresponds to orbital decay, which also assumes a circular orbit but incorporates an additional quadratic term, $\frac{dP}{dE}$, representing a steady variation of the orbital period with epoch.  

In general, detecting orbital decay in an exoplanetary system requires a long observational baseline, typically exceeding a decade. By compiling new transit times in the present study, we extend the temporal coverage of the dataset, thereby improving its suitability for investigating potential orbital decay. The orbital decay model is expressed as:
\begin{equation}
T_{q} (E) = {T}_{0} + P E + \ \frac{1}{2} \ \frac{dP}{dE} \ {E}^2,
\label{lab:equation}
\end{equation}
where $\frac{dP}{dE}$ denotes the rate of change of the orbital period $P$.

For the orbital-decay model, we followed the same fitting procedure as for the linear model, but ran 32\,000 MCMC steps per walker to sample the posterior distribution and determine the best-fit ephemeris and uncertainties for the three free parameters: $P$, $T_0$, and $\mathrm{d}P/\mathrm{d}E$. The initial 60 steps (approximately twice the integrated autocorrelation time) were discarded as burn-in to mitigate the effects of correlated samples \citep{2016A&A...595L...5A}. The prior on the period derivative, $\mathrm{d}P/\mathrm{d}E$, was allowed to vary freely over both positive and negative values.

By substituting the derived values of $P$ and $\frac{dP}{dE}$ (see Table~\ref{tb:timing_models WASP-12b}) for WASP-12 b into equation~(4) of \citet{2017AJ....154....4P},  

\begin{equation}
    \frac{dP}{dt} = \frac{1}{P} \frac{dP}{dE},
\end{equation}

we obtain the orbital decay rate as $\dot{P} \approx -31.97 \pm 0.80~\mathrm{ms~yr^{-1}}$. This result provides strong evidence that the orbit of WASP-12 b is decaying rapidly, as inferred from the full set of available transit-timing measurements spanning a baseline of $\sim 15$ years.

The third model considered is the \textit{apsidal precession model}, which assumes that the planet's orbit is slightly eccentric and that the argument of periastron ($\omega$) undergoes uniform precession. The analytical form of this model, originally proposed by \citet{1995Ap&SS.226...99G}, is given by:

\begin{equation}
    T_{ap}(E) = T_{ap0} + P_s E - \frac{e P_s}{\pi} \left[ \cos \left( \omega_0 + \frac{d\omega}{dE} E \right) - \cos \omega_0 \right],
\end{equation}

where the free parameters are defined as follows: $T_{ap0}$ is the mid-transit time at $E = 0$, $P_s$ is the sidereal period, $e$ is the orbital eccentricity, $\omega_0$ is the argument of periastron at the reference epoch ($E = 0$), and $\frac{d\omega}{dE}$ is the precession rate of the periastron. In this context, $E$ denotes the epoch, $T_{ap}(E)$ represents the calculated mid-transit time at epoch $E$, and $\omega$ corresponds to the angle between the ascending node in the plane of the sky and the orbital periastron. To infer the best-fit ephemeris of the apsidal precession model, we followed the same fitting approach outlined for linear model fitting and orbital decay model fitting, while extending the MCMC sampling to $10^{5}$ steps per walker.  

\begin{figure}
    \centering
    \includegraphics[width=1.1\linewidth]{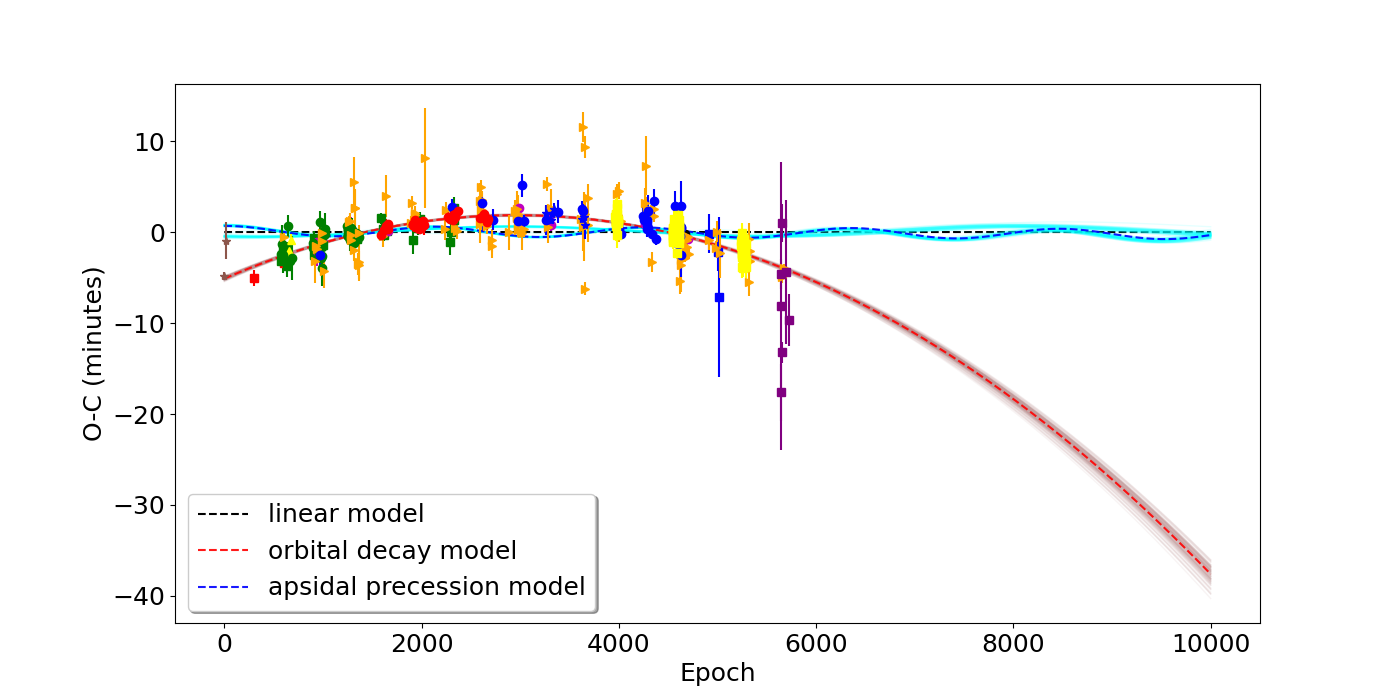}
    \caption{O-C diagram for analysing 391 mid-transit times of WASP-12b. The blue filled square show the data from \citet{2024PSJ.....5..163A}, the red filled squares are from \citet{2011AJ....141..179C}, the green filled square is from \citet{2017AJ....153...78C},  the black filled circle is from \citet{2024MNRAS.534..800A}, the brown filled asterisks are from \citet{2009ApJ...693.1920H}, the magenta filled circles are from \citet{2024A&A...686A..84L}, the yellow filled upward triangles are from \citet{2011MNRAS.411.1204M}, the green filled circles are from \citet{2013A&A...551A.108M}, the orange filled right triangles are from the quality 1 and quality 2 data of ETD, the blue filled circles are from Exoclock, the red filled circle is from \citet{2016A&A...588L...6M}, the blue filled asterisk is from \citet{2020ApJ...888L...5Y}, the yellow filled squares are from TESS data, the purple filled squares are from the new ground-based data for 7 nights. The dashed black line, red and blue curves represent the linear, orbital decay and apsidal precession models. The lines are drawn for 100 randomly chosen sets of parameters from the Markov chains of posteriors of the orbital decay (brown) and apsidal precession (cyan) models. The models are extrapolated for the next $\sim 10 $ years to illustrate the broad spectrum of possible solutions.}
    \label{fig:O-C_diagram WASP-12b}
\end{figure}

The best-fitting ephemerides derived from fitting the linear, orbital-decay, and apsidal-precession models are summarized in Table~\ref{tb:timing_models WASP-12b}. By analyzing the timing residuals obtained after fitting the three different timing models, we construct the observed minus calculated (O-C) diagram, as shown in Figure~ \ref{fig:O-C_diagram WASP-12b}. The O-C diagram serves as a powerful diagnostic tool for detecting long-term variations in the orbital period. In this case, the orbital decay model, represented by the red dashed line, reveals a clear downward trend. 
Furthermore, we randomly drew a sample of 100 parameter sets from the posterior distributions of the orbital decay model (represented by the brown solid lines in the O-C diagram) and extrapolated them over the next $\sim$13 years to investigate the projected evolution of the decay. The resulting curves consistently exhibit the same declining trend, confirming the robustness of the predicted orbital decay behavior. In the O-C diagram, the blue dashed curve represents the apsidal precession model.  To investigate the future behavior of WASP-12 b under apsidal precession, we applied the same procedure as for the orbital decay model. We randomly drew a sample of 100 parameter sets from the posterior distributions of the apsidal precession model (represented by the cyan line) and extrapolated them over the next $\sim$13 years. The resulting O-C diagram indicates that the apsidal precession model does not exhibit any significant deviation from the linear ephemeris.

Taken together, the evidence from the O-C diagram and the negative value of $\dot{P}$ derived from our model fitting strongly indicate that the orbit of WASP-12 b is shrinking over time. This provides compelling support for the interpretation that the planet is undergoing orbital decay as a consequence of tidal interactions with its host star.

\begin{table*}[!htb]
\caption{Best-Fit Model Parameters for WASP-12b}
    \centering
    \begin{tabular}{lcccc}
     \hline
     Parameter                          &  Symbol &   units    & Posterior value  & 1 $\sigma$ uncertainty  \\
     \hline
     \multicolumn{5}{c}{\textbf{Constant Period Model}} \\
        Period                          & P$_{\text{orb}}$          & days         & 1.0914187 & $^{+0.157\times 10^{-7}}_{-0.157\times 10^{-7}}$    \\
        Mid-transit time                & T$_{0}$   &BJD$_{\rm TDB}$     & 2454515.52882715  &$^{+0.494 \times 10^{-4}}_{-0.495 \times 10^{-4}}$ \\
        \hline                                                              
        N$_{dof}$                         &            &              &389         & \\
        $\chi^{2}$, $\chi^{2}_{red}$	                    &            &              & 2431.25, 6.25       &       \\
        AIC                             &           &               & 2433.64 \\
        BIC                             &           &               & 2441.57 \\
     \hline
     \hline 
    \multicolumn{5}{c}{ \textbf{Orbital Decay Model}}\\  
        Period                         & P$_{\text{orb}}$        &  days         & 1.0914220 & $^{+8.36 \times 10^{-8}}_{-8.33 \times 10^{-8}}$ \\
        Mid-transit time               & T$_{q0}$    &BJD$_{\rm TDB}$      & 2454515.52526934 & $^{+0.10 \times 10^{-3}}_{-0.10 \times 10^{-3}}$ \\
        Decay Rate                      & dP/dE     &days/epoch     & -1.10577512 $\times$ 10$^{-9}$ &  $^{+0.28\times10^{-10}}_{-0.28\times10^{-10}}$  \\
        Decay Rate                      & dP/dt     & ms/yr       & -31.97                   & 0.80      \\
        \hline
        N$_{dof}$                         &            &              &388        & \\
        $\chi^{2}$, $\chi^{2}_{red}$	                    &            &              & 865.24, 2.23       &       \\
        AIC                             &           &               & 871.68 \\
        BIC                             &           &               & 883.58 \\
        \hline 
        \hline
       \multicolumn{5}{c}{  \textbf{Apsidal Precession Model}} \\
        Sidereal Period                  & P$_{s}$    & days          &  1.09141868      &  $^{+3.64 \times 10^{-8}}_{-2.17 \times 10^{-8}}$             \\
        Mid-transit time                 & T$_{ap0}$    & BJD$_{\rm TDB}$     & 2454515.52891730    & $^{+6.19 \times 10^{-5}}_{-7.48 \times 10^{-5}}$                  \\
        Eccentricity                    & e         &               &  0.0030         & $^{+3.24 \times 10^{-5}}_{-1.75 \times 10^{-3}}$ \\
        Argument of Periastron          & $\omega$$_{0}$& rad           &    3.10          &$^{+0.037}_{-3.56}$               \\
        Precession Rate                 & d$\omega$/dE& rad/epoch     &  0.0011          & $^{+0.59 \times 10^{-5}}_{-1.78 \times 10^{-5}}$                 \\
       \hline                                                          
        N$_{dof}$                         &            &              &386         & \\
        $\chi^{2}$, $\chi^{2}_{red}$	                    &            &              & 
        2183.76, 5.66       &       \\
        AIC                             &           &               & 2194.10 \\
        BIC                             &           &               & 2213.94 \\
        \hline 
        \hline
    \end{tabular}
   
    \label{tb:timing_models WASP-12b}
\end{table*}

\subsection{Goodness-of-fit metrics}

To assess the goodness of fit, we calculated the reduced chi-square value 
($\chi^{2}$ per degree of freedom, denoted as $\chi^{2}_{r}$) for the three 
best-fit models (Table~\ref{tb:timing_models WASP-12b}) using the expression  ${\chi}^{2}_{r}$  = ${\chi}^{2}$/n, where $n$ is the number of degrees of freedom. Among the tested models, the 
orbital decay model yields the lowest reduced chi-square value 
($\chi^{2}_{r} = 2.23$ with 388 degrees of freedom) compared to the linear 
model ($\chi^{2}_{r} = 6.25$ with 389 degrees of freedom) and the apsidal 
precession model ($\chi^{2}_{r} = 5.66$ with 386 degrees of freedom). This 
indicates that the orbital-decay model provides a superior fit to the transit 
timing data of WASP-12 b.  

To further evaluate the statistical preference among the models, we employed two 
widely used information criteria: the Akaike Information Criterion (AIC; 
\citealt{1974ITAC...19..716A}) and the Bayesian Information Criterion (BIC; 
\citealt{1978AnSta...6..461S}). These are defined as  $\mathrm{AIC} = \chi^{2} + 2k_{F}$, $\mathrm{BIC} = \chi^{2} + k_{F} \ln N_{P}$, where $k_{F}$ is the number of free parameters in the model, and $N_{P}$ is the total number of data points ($N_{P} = 391$ in this analysis). For the models 
considered here, $k_{F} = 2$ for the linear ephemeris, $k_{F} = 3$ for the orbital 
decay model, and $k_{F} = 5$ for the apsidal precession model. The corresponding 
values of $\chi^{2}$, AIC, and BIC are reported in Table~\ref{tb:timing_models WASP-12b}.  

To determine whether the linear or quadratic ephemeris is favored, we calculated 
the BIC difference,  

\begin{equation}
\Delta \mathrm{BIC} = \mathrm{BIC}_{\mathrm{lin}} - \mathrm{BIC}_{\mathrm{quad}} 
= 1557.99.
\end{equation}

Since a lower BIC value corresponds to a better model fit, a positive 
$\Delta \mathrm{BIC}$ favors the quadratic (orbital decay) model. The criterion 
$\Delta \mathrm{BIC} > 10$ indicates strong evidence in support of the quadratic 
model. Given that our $\Delta \mathrm{BIC}$ far exceeds this threshold, there is 
decisive evidence for orbital decay in WASP-12 b.  

The Bayes factor ($B$) can also be estimated from $\Delta \mathrm{BIC}$ under the 
assumption of Gaussian posteriors:  

\begin{equation}
B = \exp\left(\frac{\Delta \mathrm{BIC}}{2}\right) \approx 1.58 \times 10^{338}.
\end{equation}

This result shows that the quadratic (orbital decay) model is strongly
favored over the linear model. A similar comparison using the AIC gives,  

\begin{equation}
\Delta \mathrm{AIC} = \mathrm{AIC}_{\mathrm{lin}} - \mathrm{AIC}_{\mathrm{quad}} 
= 1165.98,
\end{equation}

which further supports orbital decay as the most plausible explanation of the 
observed transit-timing variations. These findings are consistent with previous 
studies (\citealt{Maciejewski_2016, 2017AJ....154....4P, 2020ApJ...888L...5Y}).

\section{Discussions}
\label{Discussions}
\subsection{Comparison of the orbital decay rate with previous studies}

Earlier transit-timing studies (\citealt{2017AJ....154....4P, 2018AcA....68..371M, 2021AJ....161...72T}) 
reported compelling evidence for a decreasing orbital period. By incorporating a 
substantial set of newly available transit timings from ETD and ExoClock, along with 
additional ground-based photometric observations that significantly extend the temporal 
baseline, we performed an updated ephemeris analysis to obtain a more precise estimate 
of the orbital decay rate.  

Our derived value of the orbital period derivative is  

\begin{equation}
\dot{P} = -31.97 \pm 0.80~\mathrm{ms\,yr^{-1}},
\end{equation}

which is fully consistent with recent measurements and lies within the $1\sigma$ confidence interval of several reported values.  
For instance, \citet{2025arXiv250818355S}, who included the recent TESS sectors together with the IW22 data, reported  $\dot{P} = -30.85 \pm 0.82~\mathrm{ms\,yr^{-1}}$.  Similarly, \citet{2024A&A...686A..84L}, incorporating both spectroscopic and photometric data, obtained  $\dot{P} = -30.72 \pm 2.67~\mathrm{ms\,yr^{-1}}$.  Using only TESS data,\citet{2022ApJS..259...62I} reported  $\dot{P} = -30.27 \pm 1.11~\mathrm{ms\,yr^{-1}}$.  

Our derived value is also in agreement with earlier results, lying within $\sim1.4\sigma$ of the estimate  $\dot{P} = -2.56 \pm 4.0~\mathrm{ms\,yr^{-1}}$ reported by \citet{2016A&A...588L...6M}.  Furthermore, it remains consistent with the results of \citet{2017AJ....154....4P} and \citet{2020ApJ...888L...5Y}, who included occultation data along with transit data in their analyses.

A comparison of our result with values derived in earlier studies is summarized in 
Table~\ref{tab:Comparison of orbital decay rate WASP-12b}.

\begin{table*}[h]
    \centering\small\centering\renewcommand{\arraystretch}{1.4}
        \caption{Comparison of the values of period change rate of WASP-12b as estimated by previous studies.}
        \label{tab:Comparison of orbital decay rate WASP-12b}
    \begin{tabular}{l c}
    \hline\hline
         Reference & Period change rate, {$\dot {P}$} \\
         & [ms/yr] \\
         \hline
         This work & $-$31.97 $\pm$ 0.80 \\
         \citet{2025arXiv250818355S} & $-$30.85 $\pm$ 0.82 \\ 
         \citet{2024RNAAS...8..223S} & $-$26.31 $\pm$ 0.90 \\ 
        \citet{Leonardi_2024} & $-$30.72 $\pm$ 2.67 \\ 
         \citet{2024PSJ.....5..163A} & $-$29.8 $\pm$ 1.6 \\ 
         \citet{2024MNRAS.534..800A} & $-$29.5 $\pm$ 1.0 \\
         \citet{2022ApJS..259...62I} & $-$30.27 $\pm$ 1.11  \\
         \citet{2022AJ....163..175W} & $-$29.81 $\pm$ 0.94   \\
         \citet{2022MNRAS.512.3113B} & $-$37.14 $\pm$ 1.31   \\
          \citet{2021AJ....161...72T} & $-$32.53 $\pm$ 1.62  \\
         \citet{2020ApJ...888L...5Y} &  $-$29 $\pm$ 2 \\
         \citet{2017AJ....154....4P} & $-$29 $\pm$ 3  \\ 
         \citet{Maciejewski_2016} & $-$25.6 $\pm$ 4.0  \\ 
         \hline
    \end{tabular}
    
    \label{tab:comparison_of_orbital_decay_rate}

\end{table*}

\subsection{Calculation of Orbital Decay Timescale of WASP-12 b}

The orbital decay timescale ($T_d$) quantifies the characteristic time over which a hot Jupiter gradually spirals inward due to tidal interactions with its host star, ultimately leading to its engulfment. This timescale provides an important measure of the long-term dynamical evolution of close-in giant planets. Using the orbital period ($P$) and the orbital decay rate ($\dot{P}$) derived in Section~\ref{Transit timing analysis WASP-12b}, we estimate the decay timescale as

\begin{equation}
    T_d = \frac{P}{\dot{P}} \approx 2.94 \ \text{Myr}.
\end{equation}

Our derived value of $T_d$ is fully consistent with $T_d = 2.90$~Myr reported by \citet{2021AJ....161...72T}, and also with $T_d = 3.25$~Myr obtained by \citet{2020ApJ...888L...5Y}. This close correspondence reinforces the reliability of the measured orbital decay rate and suggests a consistently rapid inward migration for WASP-12 b over astrophysical timescales.

\subsection{Calculation of Stellar Tidal Quality Factor}
\label{Stellar tidal quality factor of WASP-12 b}

The identification of close-in hot Jupiters has renewed focus on the role of tidal interactions in governing stellar energy dissipation (\citealt{2018ARA&A..56..175D}). Central to this process is the tidal quality factor, ${Q}{\ast}$, which serves as a measure of how effectively a star dissipates tidal energy. Formally, ${Q}{\ast}$ is expressed as the ratio between the maximum energy stored in the tidal distortion of the star during an orbital cycle and the total energy lost to frictional dissipation in the same interval (see, e.g., Eq. (2.19) of \citealt{2008EAS....29...67Z}). This quantity is fundamental in determining the characteristic timescales over which star-planet tidal interactions affect the stellar spin evolution and the orbital dynamics of the companion. 

Since WASP-12b exhibits a markedly negative orbital decay rate (see Section~\ref{Transit timing analysis WASP-12b}), the reduction in its orbital period is most plausibly explained by tidal dissipation within the host star. On this basis, we determine the modified stellar tidal quality factor ($Q'_{\ast}$), a dimensionless parameter that empirically characterizes the efficiency of tidal kinetic energy dissipation within the star. To compute $Q'_{\star}$ for the WASP-12 system, we adopt the formalism of the 
modified constant phase-lag model (\citealt{1966Icar....5..375G}), following the methodology 
of \citet{2017AJ....154....4P}, \citet{2017ApJ...836L..24W}, and \citet{2018AcA....68..371M}. The relevant 
expression (Equation~7) is given by  

\begin{equation}
{Q}^{'}_{\ast} = -\frac{27}{2}{\pi}\left(\frac{M_p}{M_\ast}\right)\left(\frac{a}{R_\ast}\right)^{-5}\left(\frac{1}{\dot{P}}\right),
\label{eqn:quality_factor}
\end{equation}

where $P$ is the orbital period derived from the decay model, $\dot{P}$ is the 
measured orbital decay rate, $M_{p}/M_{\star}$ is the planet-to-star mass ratio, 
and $a/R_{\star}$ is the ratio of orbital semi-major axis to stellar radius, 
under the assumption that the stellar spin frequency is much smaller than the 
planetary orbital frequency.  

Adopting $M_{p}/M_{\star}$ and $a/R_{\star} = 3.061$ from 
\citet{2022AJ....163..175W}, and substituting the measured $\dot{P}$ (from Section~\ref{Transit timing analysis WASP-12b}), we infer $Q'_{\star} \approx (1.52 \pm 0.038) \times 10^{5}$ for WASP-12. This value lies within the range previously inferred for hot-Jupiter host stars ($10^{5}$–$10^{6.5}$; \citealt{2008ApJ...681.1631J, 2012MNRAS.422.3151H, 2020MNRAS.498.2270B}), for binary systems ($10^{5}$–$10^{7}$; \citealt{2005ApJ...620..970M, 2007ApJ...661.1180O, 2010A&A...512A..77L, 2015Natur.517..589M}), and for transiting giant planets ($10^{4}$–$10^{8}$; \citealt{2017A&A...602A.107B}). This value is also consistent with estimates for stars hosting gas-giant planets on ultra-short orbital periods, which typically exhibit $Q'_{\star}$ values in the range of $10^{5}$–$10^{7}$ (\citealt{2018AJ....155..165P}). Furthermore, the modified stellar tidal quality factor ($Q'_{\star}$) obtained in our analysis is of the same order of magnitude as $Q'_{\star} = 4.3 \times 10^{5}$, as derived from the theoretical models of \citet{2016ApJ...816...18E} for solar-type host stars.

While our derived $Q'_{\star}$ agrees with these results, it is 1–2 orders of 
magnitude smaller than the typical values reported for Sun-like primaries in 
eclipsing binaries ($\sim 10^{7.8}$; \citealt{2022MNRAS.512.3651P}), simplified tidal 
evolution models ($10^{7.5}$–$10^{8.5}$; \citealt{2010ApJ...723..285H}), and hot Jupiters 
in dynamical and equilibrium tide regimes ($10^{7.3}$–$10^{8.3}$; 
\citealt{2018MNRAS.476.2542C}). Our estimate of $Q'_{\star} = 1.6 \times 10^{5}$ therefore 
implies efficient tidal dissipation and rapid orbital decay. Moreover, this result 
falls within the $1\sigma$ lower confidence bound of the value reported by 
\citet{2025arXiv250818355S}, namely $Q'_{\star} = 1.64 \times 10^{5}$. A comparison of 
our result with values from earlier studies is summarized in Table~\ref{tab:comparison_of_tidal_quality_factor WASP-12b}.  

\begin{table*}[htbp]
    \centering
    \small
    \renewcommand{\arraystretch}{1.4}
        \caption{Comparison of the values of Stellar Tidal Quality Factor of WASP-12 b as estimated by previous works.}
    \begin{tabular}{l c}
    \hline\hline
         Stellar tidal quality factor, ${Q}^{'}_{\ast}$ & Reference \\
        
         \hline
         $(1.52 \pm 0.038)\times 10^{5}$ & This Work\\
          $(1.64 \pm 0.04)\times 10^{5}$ & \citet{2025arXiv250818355S}\\
           $(2.13 \pm 0.18)\times 10^{5}$ & \citet{Leonardi_2024}\\
          $(~ 1.7 \pm 0.0405\times 10^{5}$ & \citet{2024NewA..10602130Y}\\
          $(1.7 \pm 0.14)\times 10^{5}$ & \citet{akinsanmi2024tidaldeformationatmospherewasp12b}\\
         $(1.6 \pm 0.1)\times 10^{5}$ & \citet{2024MNRAS.534..800A}\\
          $(1.5 \pm 0.11)\times 10^{5}$ & \citet{2022AJ....163..175W}\\
            $(1.39 \pm 0.15)\times 10^{5}$ & \citet{2021AJ....161...72T}\\
            $(1.75 \pm 0.12)\times 10^{5}$ & \citet{2020ApJ...888L...5Y}\\
         $2.5 \times 10^{5}$ & \citet{Maciejewski_2016}\textsuperscript{a}\\
         $4.3 \times 10^{5}$ & \citet{2016ApJ...816...18E}\textsuperscript{a}\\
         \hline
         \hline
    \end{tabular}
    
    \label{tab:comparison_of_tidal_quality_factor WASP-12b}
    {Notes:
$^a$ It does not report the uncertainty (error bar) associated with the modified stellar tidal quality factor in its respective paper.}\\
\end{table*}

\subsection{Calculation of the shift in transit time and the Remaining Lifetime}

The expected shift in the transit arrival time of an exoplanet, denoted as 
$T_{\mathrm{shift}}$, corresponds to the predicted variation in transit timing 
as observed from Earth, arising from perturbations in the planet's orbital 
dynamics. To estimate the anticipated timing shift for WASP-12 b under the 
influence of orbital decay, we employed Equation~(7) of \citet{2014MNRAS.440.1470B}:

\begin{equation}
T_{\rm shift}=\frac{1}{2}T^{2}\left(\frac{dn}{dT}\right)\left(\frac{P}{2\pi}\right),
\label{eqn:shift}
\end{equation}

where $dn/dT$ represents the present rate of change in the orbital frequency 
of the planet. The orbital period ($P$) was adopted for the orbital decay model from the Table~\ref{tb:timing_models WASP-12b} . Using the calculated value $dn/dT \approx 7.16 \times 10^{-19}~\mathrm{rad\,s^{-2}}$, corresponding to a modified stellar tidal quality factor of 
$Q'_{\star} = 1.52 \times 10^{5}$, the predicted transit timing shift after 
$T = 17$~yr is found to be 
$T_{\mathrm{shift}} \approx 1546.02$~s. This prediction can be tested and 
refined through future follow-up transit monitoring.  

The remaining lifetime of a hot Jupiter corresponds to the timescale over 
which its orbit decays sufficiently for the planet to spiral inward and 
eventually merge with its host star. This orbital decay is predominantly 
governed by stellar tidal dissipation, commonly characterized by the modified 
stellar tidal quality factor ($Q'_{\star}$), which quantifies the efficiency 
of tidal energy dissipation in the star. Consequently, the orbital evolution 
timescale of hot Jupiters is strongly dependent on the adopted value of 
$Q'_{\star}$. By substituting the appropriate $Q'_{\star}$ value (see Section~\ref{Stellar tidal quality factor of WASP-12 b}) along with other system  parameters from \citet{2022AJ....163..175W} into Equation~(5) of \citet{2009ApJ...692L...9L}:

\begin{equation}
T_{\text{remain}} = \frac{1}{48}\frac{Q'_{\star}}{n}\,\left(\frac{a}{R_{\star}}\right)^5\,\left(\frac{M_{\star}}{M_p}\right) \,,
\label{timing_a}
\end{equation}

we estimate the remaining lifetime of WASP-12 b to be 
$\sim 0.41$~Myr, where $n = 2\pi/P$ denotes the mean orbital motion of the planet.

\subsection{Estimation of Planetary Love Number}

Tidal evolution theory predicts that hot Jupiters should  circularize on timescales much shorter than the ages of their host stars  (\citealt{2007A&A...462L...5L, 2018ARA&A..56..175D}). 
Assuming a planetary tidal dissipation factor of $Q_{p} \sim 10^{6}$ and applying the formalism of \citet{2017AJ....154....4P}, we estimate the tidal circularization timescale of WASP-12 b to be $\sim$ 0.34 ~Myr. This timescale is several orders of magnitude shorter than the estimated stellar age (WASP-12: $\sim$3.05 ~Gyr, \citealt{Leonardi_2024}), indicating that the planet's orbit should have long been circularized. Any residual orbital eccentricity would therefore require additional mechanisms, such as ongoing perturbations or a re-evaluation of tidal dissipation efficiency, to be sustained (\citealt{2019AJ....157..217B, 2021A&A...656A..88M}).

A plausible explanation is that apsidal precession is influenced by the planet's  internal structure, as proposed by \citet{2009ApJ...698.1778R}, since the precession  rate depends on the tidal Love number ($k_{p}$), which encodes information  about the planet's internal density distribution. Following the formulation of  \citet{2017AJ....154....4P}:
\begin{equation}
\frac{d\omega}{dE}={15}{\pi}{k_p}\left(\frac{M_\ast}{M_p}\right)\left(\frac{R_p}{a}\right)^5,
\end{equation}
we estimated $k_{p}$ by substituting the values of  $d\omega/dE$ (Table~8) together with other system parameters from  \citet{2022AJ....163..175W}. This yields  $k_{p} = 0.63 \pm 0.089$, a value consistent with Jupiter's tidal Love number  ($k_{p} = 0.59$; \citealt{2016ApJ...831...14W}). This similarity suggests that the interior  density profile of WASP-12 b may be comparable to that of Jupiter. To test this hypothesis, continued monitoring of future transits  and occultations will be essential.

\section{Conclusions}
\label{Conclusions}

Previous studies of WASP-12\,b \citep{2016A&A...588L...6M, 2017AJ....154....4P, 2020ApJ...888L...5Y} have revealed intriguing aspects of the system's orbital dynamics. Early investigations reported the presence of short-term transit timing deviations potentially caused by an additional body \citep{2011A&A...528A..65M}, while later studies identified evidence of long-term timing deviations. Specifically, \citet{2016A&A...588L...6M, 2017AJ....154....4P, 2018AcA....68..371M} observed a declining trend in transit timings indicative of orbital decay, whereas \citet{2019MNRAS.482.1872B} proposed apsidal precession as a plausible mechanism. More recently, \citet{2020ApJ...888L...5Y} provided strong evidence supporting orbital decay in the WASP-12\,b system. Detecting such orbital evolution requires long-term, high-precision monitoring. In this work, we combined newly acquired ground-based photometric data from 7 nights, obtained at various observatories worldwide, with publicly available light curves from the literature, ETD, ExoClock, and high-cadence, high-precision space-based observations from TESS to construct an extensive dataset for re-examining this hot Jupiter.

We included a total of 119 complete transit light curves of WASP-12 b observed by TESS across six sectors (20, 43, 44, 45, 71, and 72). To extend the observational time baseline, we incorporated an additional 131 high-quality transit light curves from public databases, including ETD and ExoClock, along with 108 publicly available complete transit observations. Furthermore, 26 transit light curves were obtained via private communication with the respective authors and we also included 7 new ground based observations in this study. In total, 391 light curves were modeled to refine the system's physical and orbital parameters and to determine individual mid-transit times. To ensure uniformity and precision, all light curves were analyzed using a consistent methodology for deriving the mid-transit times.

Our transit-timing analysis indicates that the orbit of WASP-12 b is decaying at a rate of $-31.97 \pm 0.80~\mathrm{ms\,yr^{-1}}$ corresponding to a modified stellar tidal quality factor of $Q'_{\star} = (1.52 \pm 0.038) \times 10^{5}$. Statistical diagnostics, including reduced $\chi^2$, BIC, and AIC, favor the orbital decay model as the most plausible explanation for the observed timing variations.

To constrain the planet's internal structure, we calculated the planetary tidal Love number, obtaining $k_{p} = 0.63 \pm 0.089$. This value is consistent with Jupiter’s Love number ($k_{p} = 0.59$; \citealt{2016ApJ...831...14W}), suggesting a similar internal density distribution for WASP-12 b. Continued long-term, high-precision monitoring of future transits with forthcoming missions such as PLATO (\citealt{2025ExA....59...26R}) and ARIEL (\citealt{2022ExA....53..807M}), complemented by secondary-eclipse (occultation) observations, will be critical for identifying the physical mechanism responsible for the observed timing offsets in WASP-12 b.


\begin{acknowledgments}
We would like to thank the anonymous reviewer for his/her valuable suggestions and constructive feedback, which have substantially enhanced the quality of this paper. I.G.J.  acknowledges funding from the National Science and Technology Council (NSTC), Taiwan, through Grants NSTC 113-2112-M-007-030, NSTC 114-2112-M-007-029, and NSTC 113-2115-M-007-008. This work utilizes new photometric observations obtained with the 0.61 m VASISTHA Telescope at the Ionospheric and Earthquake Research Centre and Optical Observatory (IERCOO), Sitapur, Paschim Midnapore, operated by ICSP. We are grateful to Prof. Sandip Kumar Chakraborty for providing access to this facility and to the IERCOO staff members for their invaluable support during the observing runs. Additionally, this study incorporates observations from the 1.3 m Devasthal Fast Optical Telescope (DFOT) at the Devasthal Campus of ARIES, Nainital, and the 0.3 m AG Optical telescope at the Utah Desert Remote Observatory (UDRO) in Beryl, Utah. We thank the staff at DFOT and UDRO for their assistance. We also acknowledge Hebb Lesbie, Efrain Alavarado, and Karen Collins for sharing their transit light curves through private communication, as well as all other authors who made their published light curves publicly available, which proved to be an invaluable resource for this study.  

This work also makes use of data from the Transiting Exoplanet Survey Satellite (TESS) mission, publicly available through the Mikulski Archive for Space Telescopes (MAST). We gratefully acknowledge the TESS mission for its significant contribution to exoplanet science by providing high-precision transit photometric data. The specific observations analyzed in this study can be accessed via the TESS LCs–All Sectors page (\citealt{10.17909/t9-nmc8-f686} ), and funding for the mission is provided by NASA's Explorer Program. Furthermore, this study utilized publicly available transit light curves from the Exoplanet Transit Database (ETD) and the ExoClock Project; we thank the contributors of these initiatives for making their data accessible. Currently, all the original lightcurves of ETD are available in VarAstro Server. 

Finally, this work made use of the NASA Exoplanet Archive, operated by the California Institute of Technology under contract with NASA as part of the Exoplanet Exploration Program.
\end{acknowledgments}

%


\appendix

\section{Graphical Representation of individual transit events taken from TESS, ETD and Exoclock} \label{app:individual_transits}
\begin{figure}
    \centering
    \includegraphics[width=1.1\linewidth]{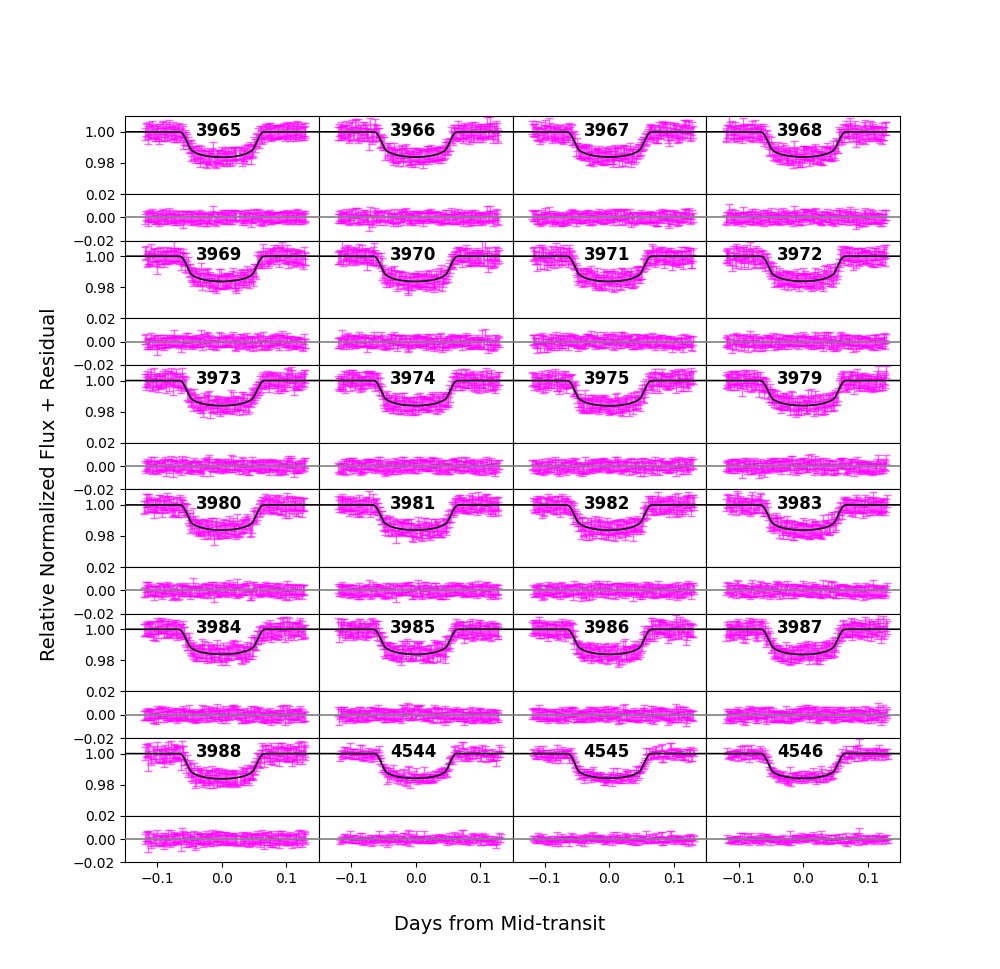}
    \caption{The upper panel shows the relative normalized flux of WASP-12 b as a function of time (expressed as the offset from the mid-transit time in TDB-based BJD), together with the epoch number of individual transits observed by TESS between epochs 3965 and 4546. The flux error bars are shown in magenta. The solid curve is the best-fit light curve model obtained through juliet. The lower panel displays the corresponding residuals with their error bars.}
    \label{fig:TESS1_WASP-12b}
\end{figure}

\begin{figure}
    \centering
    \includegraphics[width=1.1\linewidth]{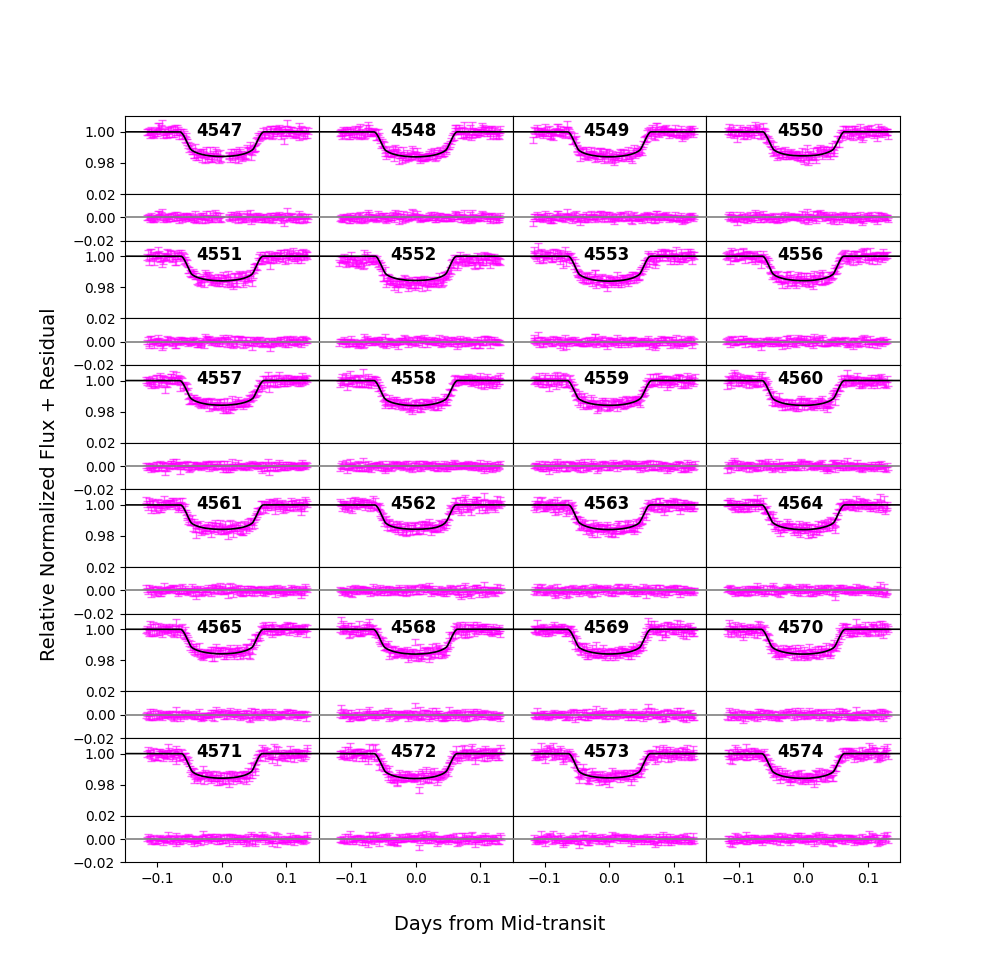}
    \caption{Same as the Figure \ref{fig:TESS1_WASP-12b} but for epochs (4547-4574)}
    \label{fig:TESS2_WASP-12b}
\end{figure}

\begin{figure}
    \centering
    \includegraphics[width=1.1\linewidth]{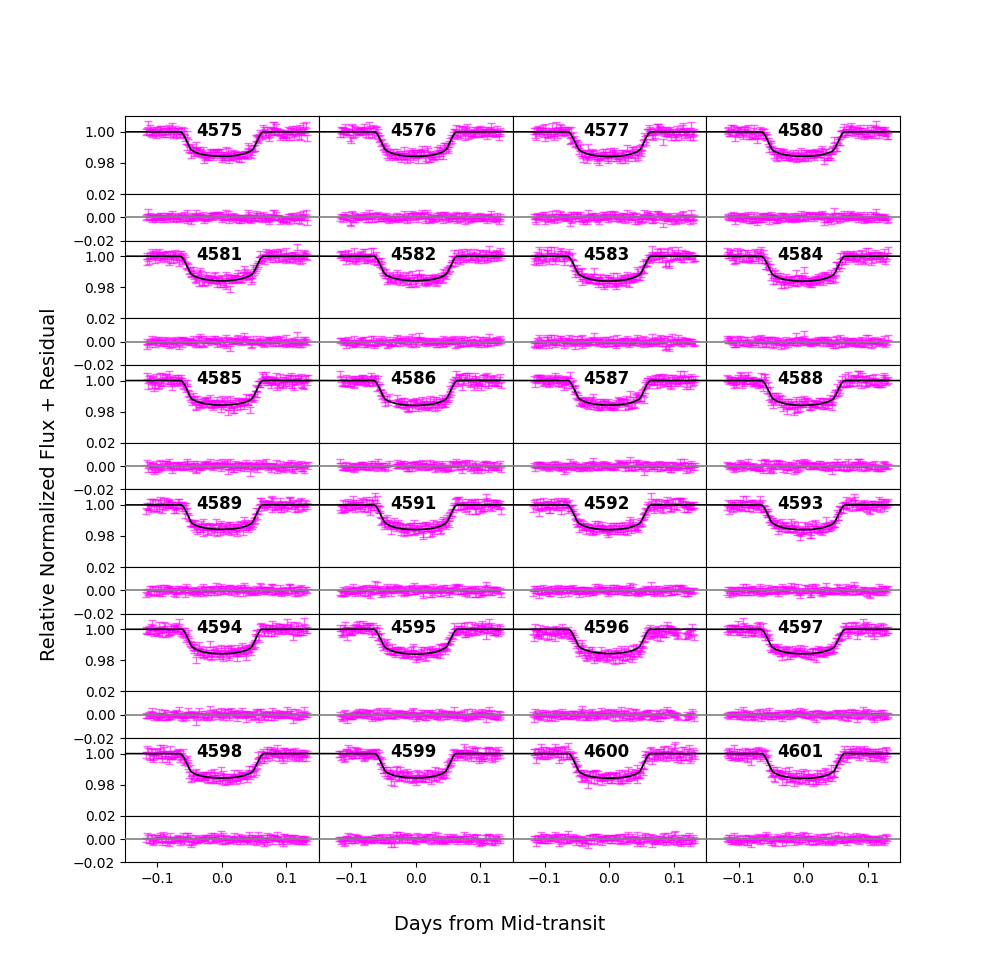}
    \caption{Same as the Figure \ref{fig:TESS1_WASP-12b} but for epochs (4575 - 4601)}
    \label{fig:TESS3_WASP-12b}
\end{figure}

\begin{figure}
    \centering
    \includegraphics[width=1.1\linewidth]{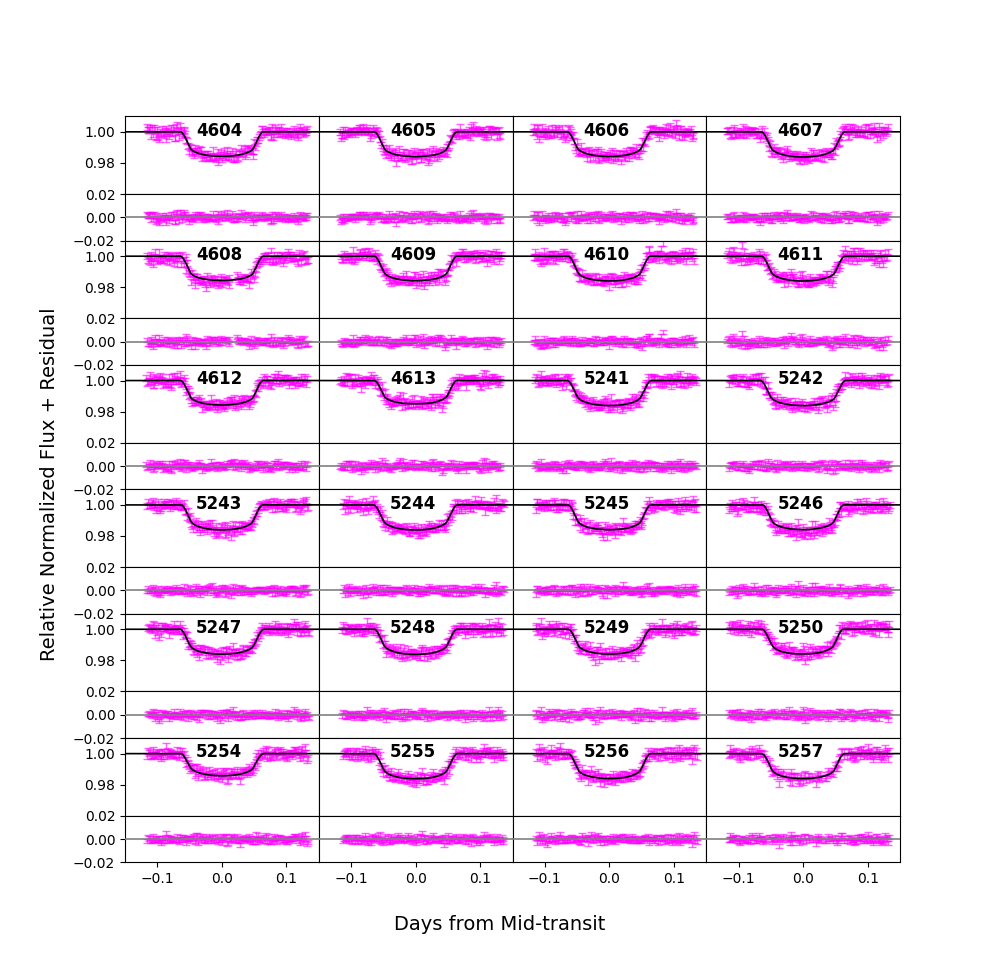}
    \caption{Same as the Figure \ref{fig:TESS1_WASP-12b} but for epochs (4604-5257)}
    \label{fig:TESS4_WASP-12b}
\end{figure}

\begin{figure}
    \centering
    \includegraphics[width=1.1\linewidth]{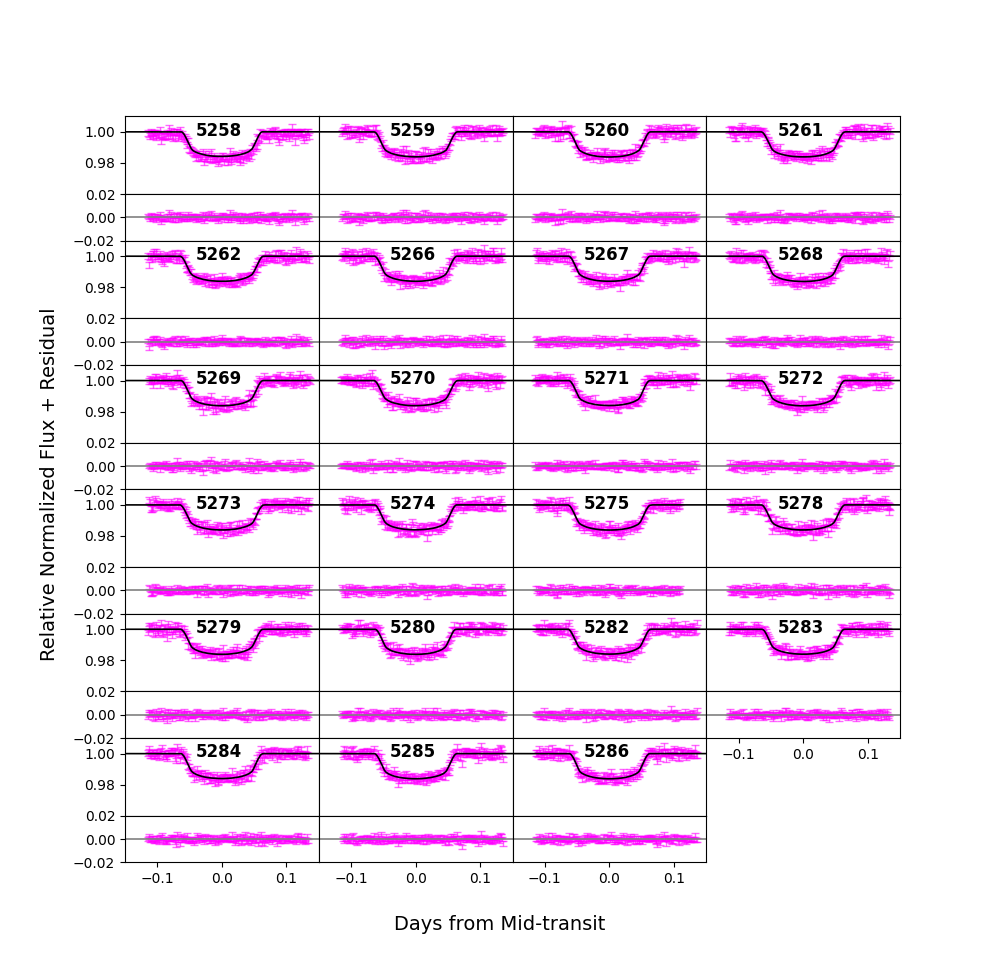}
    \caption{Same as the Figure \ref{fig:TESS1_WASP-12b} but for epochs (5258 - 5286)}
    \label{fig:TESS5_WASP-12b}
\end{figure}

\begin{figure}
    \centering
    \includegraphics[width=1.1\linewidth]{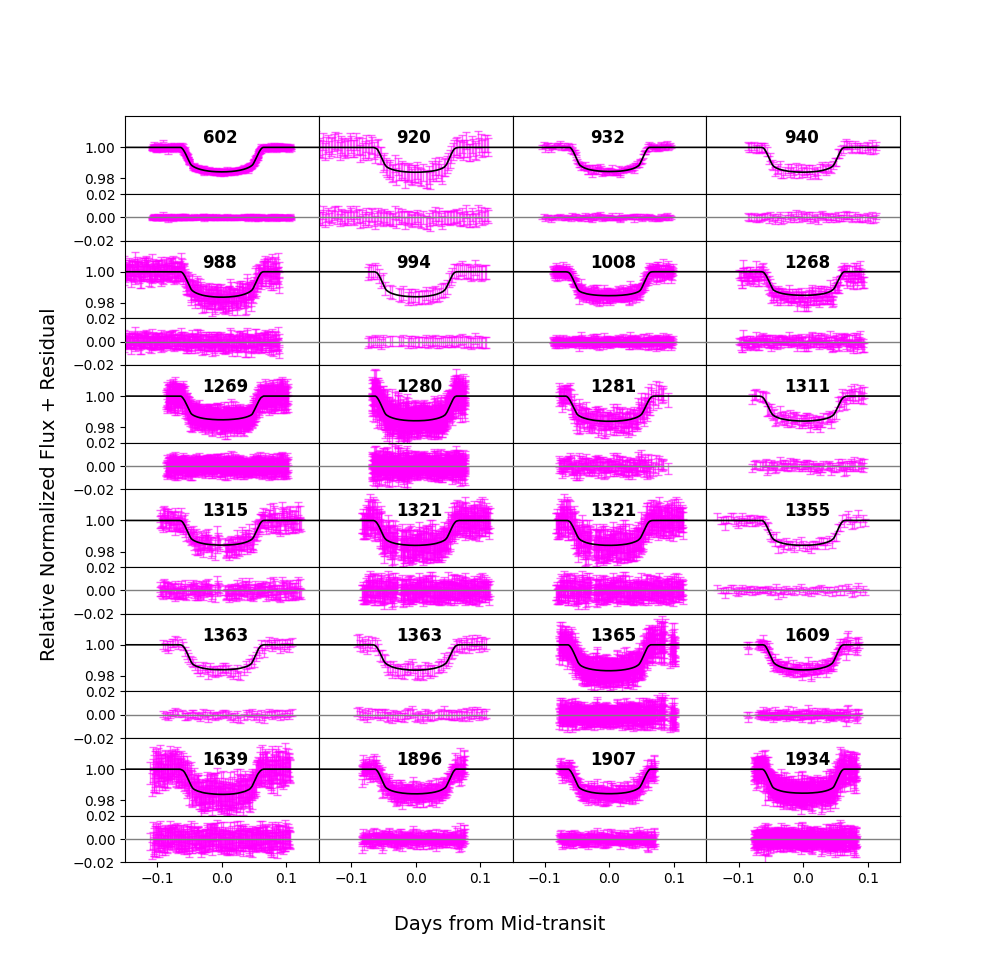}
    \caption{The upper panel shows the relative normalized flux of WASP-12 b as a function of time (expressed as the offset from the mid-transit time in TDB-based BJD), together with the epoch number of individual transits observed by ETD between epochs 602 and 1934. The flux error bars are shown in magenta and the solid curve is the best-fit light curve model obtained through juliet. The lower panel displays the corresponding residuals with their error bars.}
    \label{fig:ETD1_WASP-12b}
\end{figure}

\begin{figure}
    \centering
    \includegraphics[width=1.1\linewidth]{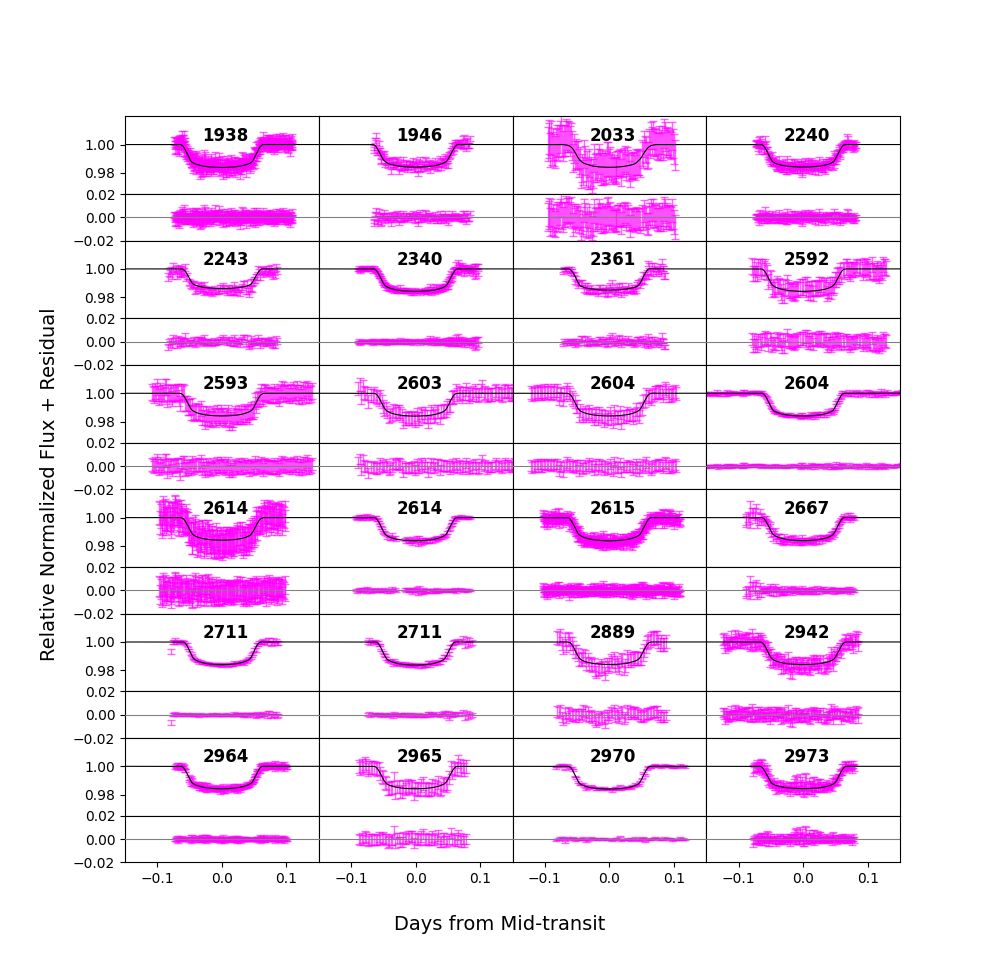}
    \caption{Same as the Figure \ref{fig:ETD1_WASP-12b} but for epochs (1938 - 2973)}
    \label{fig:ETD2_WASP-12b}
\end{figure}

\begin{figure}
    \centering
    \includegraphics[width=1.1\linewidth]{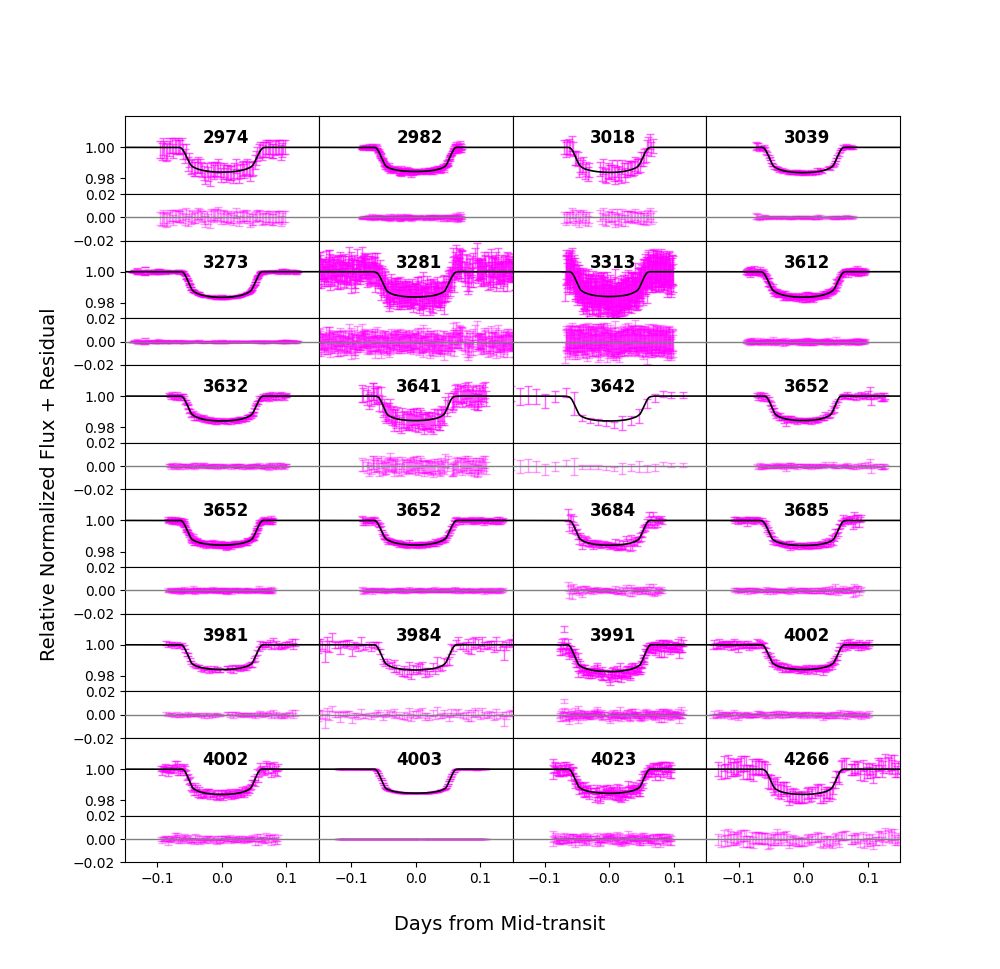}
    \caption{Same as the Figure \ref{fig:ETD1_WASP-12b} but for epochs (2974 - 4266)}
    \label{fig:ETD3_WASP-12b}
\end{figure}

\begin{figure}
    \centering
    \includegraphics[width=1.1\linewidth]{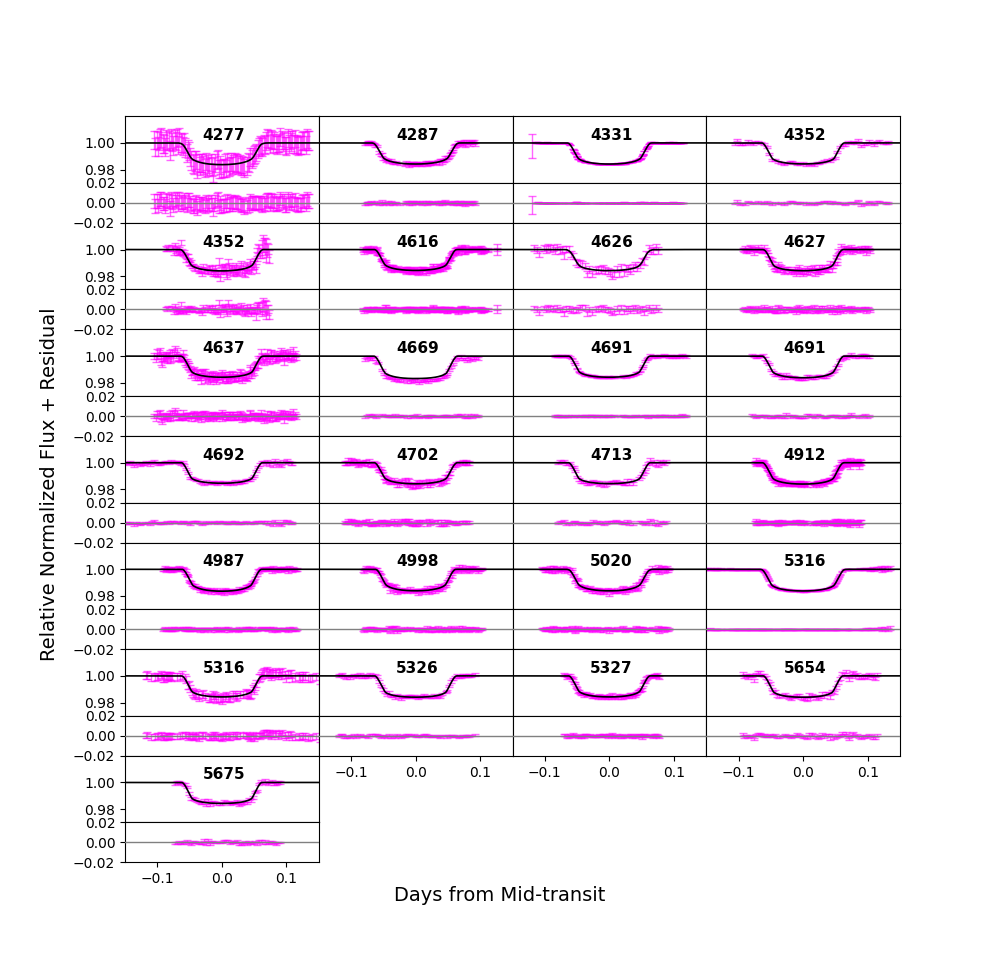}
    \caption{Same as the Figure \ref{fig:ETD1_WASP-12b} but for epochs (4277 - 5675)}
    \label{fig:ETD4_WASP-12b}
\end{figure}

\begin{figure}
    \centering
    \includegraphics[width=1.1\linewidth]{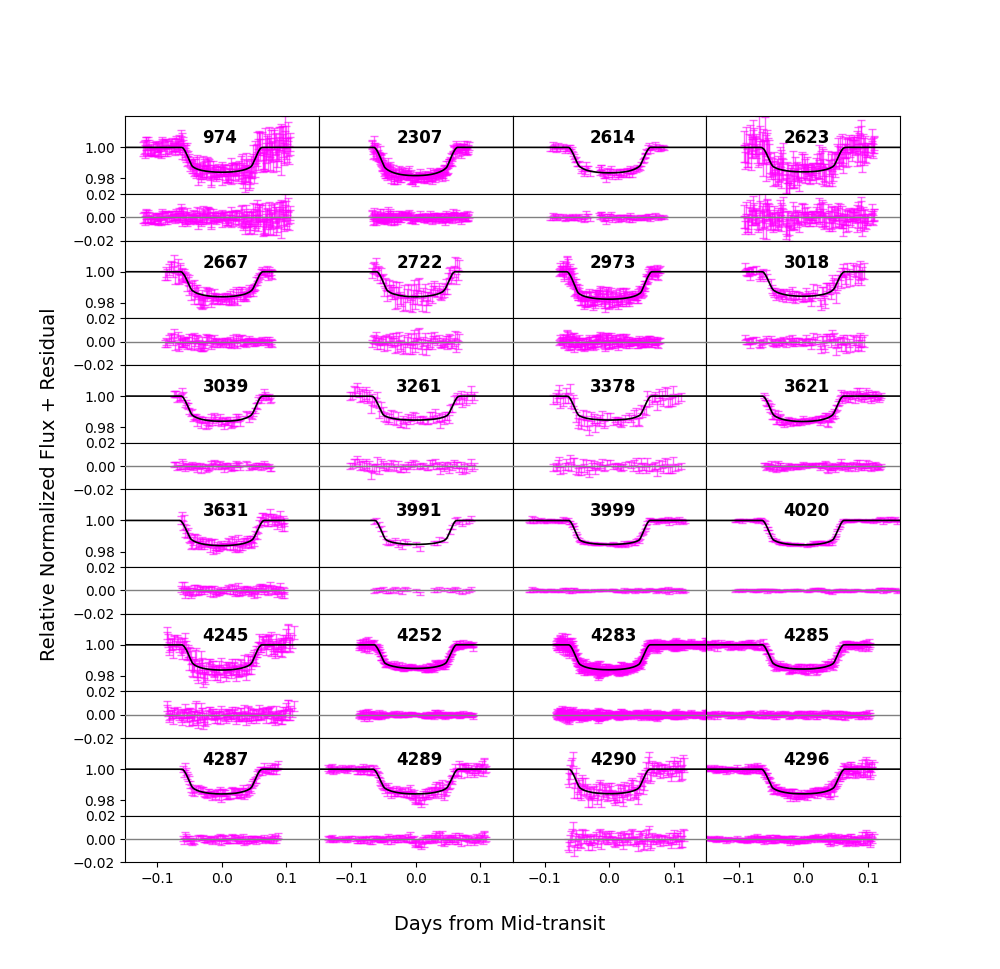}
    \caption{The upper panel shows the relative normalized flux of WASP-12 b as a function of time (expressed as the offset from the mid-transit time in TDB-based BJD), together with the epoch number of individual transits observed by Exoclock between epochs 974 and 4296. The flux error bars are shown in magenta and the solid curve is the best-fit light curve model obtained through juliet. The lower panel displays the corresponding residuals with their error bars.}
    \label{fig:Exoclock1_WASP-12b}
\end{figure}

\begin{figure}
    \centering
    \includegraphics[width=1.1\linewidth]{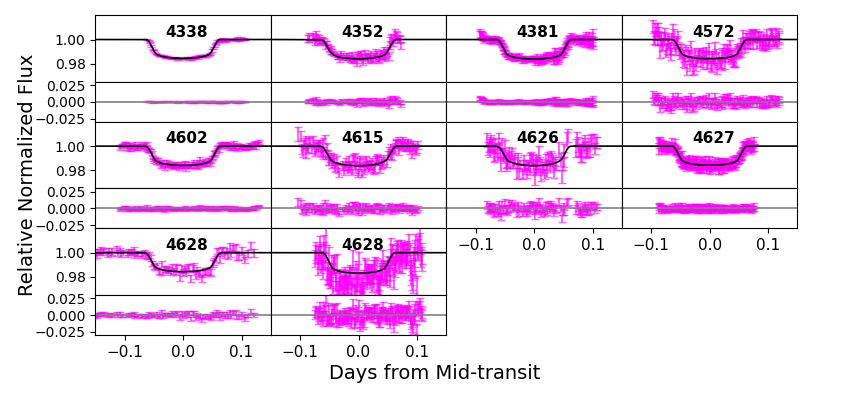}
    \caption{Same as the Figure \ref{fig:Exoclock1_WASP-12b} but for epochs (4338 - 4628)}
    \label{fig:Exoclock2_WASP-12b}
\end{figure}

We have represented model fits obtained from juliet for all  119 TESS light curves of WASP-12\,b in Figures \ref{fig:TESS1_WASP-12b}--\ref{fig:TESS5_WASP-12b}. Similarly, 97 light curves from ETD and 34 light curves from Exoclock have been represented in Figure \ref{fig:ETD1_WASP-12b}--\ref{fig:ETD4_WASP-12b} and Figure \ref{fig:Exoclock1_WASP-12b}--\ref{fig:Exoclock2_WASP-12b}.

\bibliography{wasp12b}{}
\bibliographystyle{aasjournalv7}



\end{document}